\documentclass[11pt]{article}
\usepackage{amssymb}
\usepackage{citesort}
\usepackage{url}
\usepackage{a4}
\usepackage{epsfig}

\newcommand{\ghyp}{g_{\mcH}}
\newcommand{\iniB}{B(0,\frac 12)}
\newcommand{\scrips}{{\scrip_{\tau_+}}}
\newcommand{\mcMs}{{\mcM_{\tau_+}}}
\newcommand{\bmcMs}{{\bmcM_{\tau_+}}}
\newcommand{\bhz}{\mcE_\tau^+(0)}
\newcommand{\bhzp}{\mcE_{\tau_+}^+(0)}
\newcommand{\mcOa}{\mcO_{(a,\vec 0)}}
\newcommand{\pJ}{{\dot J}}

\newcommand{\Mink}{{(\R^{3+1},\eta)}}
\newcommand{\hypo}{{\hyp_0}}
\newcommand{\hg}{{\hat g}}
\newcommand{\Sdev}{(\mcM_{\Schw},g_{\Schw})}

\newcommand{\Schw}{{\mathrm Schw}}

\newcommand{\riemgz}{g_0}

\newcommand{\del}{\partial}

\renewcommand{\hbar}{{\overline \riemgz}}

\newcommand{\const}{\mathrm{const}}

\newcommand{\mcE}{{\mycal E}}
\newcommand{\mcN}{{\mycal N}}
\newcommand{\mcA}{{\mycal A}}
\newcommand{\mcEh}{{{\mycal E}^+_\hyp}}

\newcommand{\nablash}{\nabla{\kern -.75 em
     \raise 1.5 true pt\hbox{{\bf/}}}\kern +.1 em}
\newcommand{\Deltash}{\Delta{\kern -.69 em
     \raise .2 true pt\hbox{{\bf/}}}\kern +.1 em}
\newcommand{\Rslash}{R{\kern -.60 em
     \raise 1.5 true pt\hbox{{\bf/}}}\kern +.1 em}

\newcommand{\mcO}{{\mycal O}}

\newcommand{\mcU}{{\mycal U}}
\newcommand{\mcV}{{\mycal V}}

\newcommand{\hyp}{{\mycal S}}

\newcommand{\mcH}{{\mycal H}}
\newcommand{\mcM}{{\mycal M}}
\newcommand{\bmcM}{\,\,\,\,{\widetilde{\!\!\!\!\mycal M}}}
\newcommand{\wtg}{\widetilde{g}}
\newcommand{\mcK}{{\mycal K}}

\newcommand{\bea}{\begin{eqnarray}}
\newcommand{\beaa}{\begin{eqnarray*}}
\newcommand{\bean}{\begin{eqnarray}\nonumber}

\newcommand{\bel}[1]{\begin{equation}\label{#1}}
\newcommand{\beal}[1]{\begin{eqnarray}\label{#1}}
\newcommand{\beadl}[1]{\begin{deqarr}\label{#1}}
\newcommand{\eeadl}[1]{\arrlabel{#1}\end{deqarr}}
\newcommand{\eeal}[1]{\label{#1}\end{eqnarray}}
\newcommand{\eead}[1]{\end{deqarr}}
\newcommand{\eea}{\end{eqnarray}}
\newcommand{\eeaa}{\end{eqnarray*}}

\newcommand{\be}{\begin{equation}}
\newcommand{\ee}{\end{equation}}

\newcommand{\eq}[1]{\eqref{#1}}
\newcommand{\Eq}[1]{Equation~(\ref{#1})}

\DeclareFontFamily{OT1}{rsfs}{}
\DeclareFontShape{OT1}{rsfs}{m}{n}{ <-7> rsfs5 <7-10> rsfs7 <10->
rsfs10}{} \DeclareMathAlphabet{\mycal}{OT1}{rsfs}{m}{n}
\def\scri{{\mycal I}}%
\def\scrip{\scri^{+}}%

\usepackage{amsmath} 

\let\ssection=\section

\renewcommand{\section}{\setcounter{equation}{0}\ssection}

\newtheorem{defi}{\sc Definition\rm}[section]

\newtheorem{Theorem}[defi]{\sc Theorem\rm}

\newtheorem{corollary}[defi]{\sc Corollary\rm}

\newtheorem{Proposition}[defi]{\sc Proposition\rm}

\newtheorem{Lemma}[defi]{\sc Lemma\!\rm}

\newtheorem{Remark}[defi]{{\sc Remark}\rm}

\newcommand{\qed}{\hfill $\Box$\bigskip}

\newcommand{\proof}{\noindent {\sc Proof:\ }}

\def \Reel{\mathbb{R}}

\def \R {\Reel}

\def \Nat{\mathbb{N}}

\def \N {\Nat}

\newcommand{\acknowledgments}{ \medskip \noindent{\bf Acknowledgments:} }

\newcounter{mnotecount}[section]

\renewcommand{\themnotecount}{\thesection.\arabic{mnotecount}}

\newcommand{\mnote}[1]
{\protect{\stepcounter{mnotecount}}$^{\mbox{\footnotesize $
\bullet$\themnotecount}}$ \marginpar{
\raggedright\tiny\em $\!\!\!\!\!\!\,\bullet$\themnotecount: #1} }

\newcommand{\rmnote}[1]{}

\begin{document}
\title{On ``many black hole'' vacuum space-times}
\author{
Piotr T. Chru\'sciel\thanks{Partially supported by a Polish
Research Committee grant; email \protect\url{
piotr@gargan.math.univ-tours.fr}} \\
Albert Einstein Institute\thanks{Visiting Scientist. Permanent
address: D\'epartement de math\'ematiques, Facult\'e des
Sciences, Parc de Grandmont, F37200 Tours, France.}\\
 Golm, Germany \\
\\
Rafe Mazzeo\thanks{Partially supported by NSF grant DMS-0204730;
email \protect\url{mazzeo@math.stanford.edu}} \\
Department of Mathematics \\
Stanford University\\
Stanford, CA 94305 USA}
\date{}

\maketitle


\begin{abstract} We analyze the horizon structure of families of
space times obtained by evolving initial data sets containing
apparent horizons with several connected components. We show that
under certain smallness conditions the outermost apparent horizons
will also have several connected components. We further show that,
again  under a smallness condition, the maximal globally
hyperbolic development of the many black hole initial data
constructed in \cite{ChDelay2}, or  of hyperboloidal data
of~\cite{IMP}, will have an event horizon, the intersection of
which with the initial data hypersurface is not connected. This
justifies the \emph{``many black hole"} character of those
space-times.
\end{abstract}





\section{Introduction}\label{Sintro}

 There is an
ongoing effort to construct ``many black hole" solutions of the
vacuum Einstein equations numerically (see {\em
e.g.\/}~\cite{Alcubierre:2001vm,Lehner:2001wq,Lehner:2002mn} and
references therein). In practice this means that one numerically
evolves initial data which contain trapped surfaces for as long as
the computer allows. The question then arises whether the
resulting space-time does indeed contain more than one black hole,
or for that matter, any. Several issues arise here:
\begin{itemize}
\item[a)] The notion of a black hole is usually tied to the
existence of a conformal completion of the space-time (but see
\cite{Chrusciel:2002mi} for alternative proposals). It is far from
clear that the vacuum solutions, which are in principle associated
with their numerical counterparts discussed in
\cite{Alcubierre:2001vm,Lehner:2001wq,Lehner:2002mn}, possess
sufficiently controlled conformal completions, if any.

\item[b)] Even assuming the issues in point a) do not occur,
consider an initial data set $(\hyp,g,K)$ which contains several
trapped, or marginally trapped, surfaces. If yet another trapped
or marginally trapped surface $S_0$ encloses all the previous
ones, then the geometry enclosed by $S_0$ is hidden from external
observers by the null hypersurface $\pJ ^+(S_0)$. Numerical
calculations of tentative radiation patterns inside $J^+(S_0)$
have absolutely no relevance to the data detected at $\scrip$.
Thus \emph{if} one is willing to associate black hole regions to
apparent horizons, then only the outermost apparent horizons are
relevant. In this context a  condition for a multi-black-hole
space-time would be that the {\em outermost} apparent horizon has
more than one component.

\item[c)] In any case, the event horizon itself might have nothing
to do with the apparent horizons, even the outermost ones. Under
appropriate hypotheses, the existence of an apparent horizon
implies the existence of a black hole region, but this can be much
larger than the region enclosed by the outermost apparent horizon.
In particular one might imagine a situation in which the
{outermost} apparent horizon has several components, all of which
are enclosed by a connected event horizon, so that the space-time
contains only a single black hole region.
\end{itemize}
The object of this paper is to point out that the issues raised
above can be analysed in a reasonably satisfactory way for the
``many Schwarzschild" initial data constructed in~\cite{ChDelay2},
or for the data obtained by the gluing constructions of
\cite{Joyce,IMP}, or for families of initial data sharing certain
qualitative properties, as made precise below, similar to those
of~\cite{ChDelay2,IMP}.  The first main result here is that for a
rather general class of ``small-data" families of black-hole
space-times, the outermost apparent horizon $\mcA$ will have
several connected components. We prove this both on the usual
asymptotically flat initial data hypersurfaces and on
hyperboloidal ones. For the initial data of~\cite{ChDelay2} the
relevant smallness condition holds when the mass parameters $m_i$,
$i=1,\ldots,I$, of the individual Schwarzschild black holes are
small enough as compared to the distance parameters $r_i$. For the
initial data of \cite{IMP} the smallness condition holds when the
gluing necks are sufficiently small.

One of the features of the initial data sets of~\cite{ChDelay2} is
that these metrics are \emph{exactly} Schwarzschild outside a
compact set, and this guarantees that for any one of these, the
associated maximal globally hyperbolic development $(\mcM,g)$
necessarily has a $\scrip$ which is complete to the
past.\footnote{We note that both here, and in several situations
of interest, one can use the results in~\cite{KlainermanNicolo} to
infer the existence and past-completeness of $\scrip$. However,
the estimates there are based on spherical outgoing null
hypersurfaces, which can be used to prove the existence of, at
best, a connected black hole region (if any). Further, the
differentiability properties of $\scrip$ which can be directly
inferred from that work (compare~\cite{Christodoulou:Rome}) are
not sufficient to be able to invoke the stability results
of~\cite{Friedrich}, as needed below.} As already indicated, the
existence of $\scrip$ is the usual starting point for a discussion
of black hole regions. The second main result here is the proof
that, for certain configurations and again for mass parameters
small enough, the intersection
\begin{equation}\label{0}\mcEh:=\pJ ^-(\scrip) \cap\hyp\end{equation}
of the future event horizon $\pJ ^-(\scrip)$ with the initial data
hypersurface $\hyp$ has at least $I$ components. (Indeed, we show
that $\mcA$ has \emph{exactly} $I$ components and believe that
this should also be true for $\mcEh$; a proof of such a claim
about $\mcEh$  would require complete control of the global
structure of the resulting space-time, which is well beyond the
range of techniques available nowadays.)

\section{``Many black hole'' initial data}
There are several constructions of families of initial data
containing apparent horizons,
see~\cite{Dain:2000hk,Alcubierre:1998rq,Beigneill:trapped,01705489}
and references therein. In this section we briefly describe three
such families of ``many black hole initial data". Before doing
this, it is useful to recall how apparent horizons are detected
using initial data (compare~\cite{Beigneill:trapped}): let, thus,
$(\hyp,g,K)$ be an initial data set, and let $S\subset \hyp$ be a
compact embedded two-dimensional two-sided submanifold in $\hyp$.
If $n^i$ is the field of outer normals to $S$ and $H$ is the outer
mean extrinsic curvature\footnote{We use the definition that gives
$H=2/r$ for round spheres of radius $r$ in three dimensional
Euclidean space.} of $S$ within $\hyp$ then, in a convenient
normalisation, the divergence $\theta_+$ of future directed null
geodesics normal to $S$  is given by \bel{tp} \theta_+ = H
+K_{ij}(g^{ij}-n^in^j)\;.\ee In the time-symmetric case $\theta_+$
reduces thus to $H$, and $S$ is trapped if and only if $H<0$,
marginally trapped if and only if $H=0$. In the hyperboloidal case
with $K_{ij}=g_{ij}$ we obtain $\theta_+=H+2$.

\subsection{Brill-Lindquist initial data}
Probably the simplest examples are the time-symmetric initial data
of Brill and Lindquist. Here the space-metric at time $t=0$ takes
the form
\begin{equation}\label{BL} g= \psi^4(dx^2+dy^2+dz^2)\;,\end{equation}
with $$\psi=1+\sum_{i=1}^I \frac{m_i}{2|\vec x - \vec x_i|}\;.$$
The positions of the poles $\vec x_i\in\R^3$ and the values of the
mass parameters $m_i\in \R$ are arbitrary. If all the $m_i$ are
positive and sufficiently small, then there exists a small minimal
surface with the topology of a sphere which encloses $\vec x_i$.
This follows from general arguments in geometric measure theory,
as implemented and described in more detail in Section~\ref{Sah}
below. In addition, from~\cite{KlainermanNicolo}, the associated
maximal globally hyperbolic development possesses a $\scrip$ which
is complete to the past, but the differentiability properties of
the conformally completed metric may not be sufficient to justify
 some key steps of the global analysis
below concerning event horizons.

\subsection{The ``many Schwarzschild" initial data of~\protect\cite{ChDelay2}}
\label{Sid}\rm There is a well-known special case of \eq{BL},
which is the space-part of the Schwarzschild metric centred at
$\vec x_0$ with mass $m$ : \rm
\begin{equation}\label{1}
g= \left(1 + \frac m {2|\vec x-\vec x_0|}\right )^4\delta \;,
\end{equation}
where $\delta $ is the Euclidean metric. Abusing terminology in a
standard way, we call \eq{1} simply the Schwarzschild metric.
Allowing the mass parameter to be nonpositive leads to naked
singularities or flat regions, {\em cf.}\/~\cite{ChDelay2}, but we
shall always require positive masses here. The sphere $|\vec
x-\vec x_0|=m/2$ is minimal, and the region $|\vec x-\vec x_0|<
m/2$ corresponds to the second asymptotic region, beyond the
Einstein-Rosen bridge.

Now fix the radii $0\le 4R_1<R_2<\infty$. Denoting by $B(\vec a,
R)$ the open coordinate ball centred at $\vec a$ with radius $R$,
choose points
$$
\vec x_i \in \Gamma_0(4R_1,R_2):= \begin{cases} B(0,R_2)\setminus
\overline{B(0,4R_1)}\;, & R_1 \ge 0 \cr B(0,R_2) \;, & R_1 = 0
\;,\end{cases}
$$
and radii $r_i$, $i=1,\ldots, 2N$, so that the closed balls
$\overline{B(\vec x_i,4r_i)}$ are all contained in
$\Gamma_0(4R_1,R_2)$ and are pairwise disjoint. Set
\begin{equation}\label{Om1} \Omega:= \Gamma_0(R_1,R_2) \setminus \left(\cup_i
\overline{B(\vec x_i,r_i)}\right)\;.
\end{equation}  We assume that the
$\vec x_i$ and $r_i$ are chosen so that $\Omega$ is invariant with
respect to the reflection $\vec x \to - \vec x$. Now consider a
collection of nonnegative mass parameters, arranged into a vector
as
$$\vec M=(m,m_0,m_1,\ldots ,m_{2N}),$$
where $0<2m_i < r_i$, $i\geq 1$, and in addition with $2m_0< R_1$
if $R_1 > 0$ but $m_0=0$ if $R_1 = 0$. We assume that the mass
parameters associated to the points $\vec x_i$ and $-\vec x_i$ are
the same. The remaining entry $m$ is explained below.

Given this data, it follows from the work of \cite{Corvino} (as
pointed out in~\cite{ChDelay2}, compare~\cite{ChDelay}) that there
exists a $\delta> 0$ such that if
\begin{equation}\label{mcondi}\sum_{i=0}^{2N}|m_i| \le \delta\;,
\end{equation}
then there exists a number
$$m= \sum_{i=0}^{2N}m_i + O(\delta^2)$$
and a $C^\infty$ metric $\hat g_{\vec M}$ which is a solution of
the time-symmetric vacuum constraint equation
$$R(\hat g_{\vec M})=0\;,$$ such that:
\begin{enumerate}
\item On the punctured balls $B(\vec x_i,2r_i)\setminus \{\vec
x_i\}$, $i\ge 1$, $\hg_{\vec M}$ is the Schwarzschild metric,
centred at $\vec x_i$, with mass $m_i$; \item On $\R^3 \setminus
\overline{B(0,2R_2)}$, $\hg_{\vec M}$ agrees with the
Schwarzschild metric centred at $0$, with mass $m$; \item If
$R_1>0$, then $\hat g_{\vec M}$ agrees on
$B(0,2R_1)\setminus\{0\}$ with the Schwarzschild metric centred at
$0$, with mass $m_0$.
\end{enumerate}
In fact, this construction also gives that $\hg_{\vec M}$ is
symmetric under the parity map $\vec x \to - \vec x$.


\subsection{Black holes and gluing methods}\label{Sbhgm}

A recent alternate technique for gluing initial data sets is given
in~\cite{IMP}, see also~\cite{Joyce} for the time-symmetric case
and~\cite{IMP2} for more general results in the asymptotically
Euclidean case. In this approach, general initial data sets on
compact manifolds or with asymptotically Euclidean or
hyperboloidal ends are glued together to produce solutions of the
constraint equations on the connected sum manifolds. Only very
mild restrictions on the original initial data are needed.  The
neck regions produced by this construction are again of
Schwarzschild type. The overall strategy of the construction is
similar to that used by Corvino (and in many previous gluing
constructions). Namely, one takes a family of approximate
solutions to the constraint equations and then attempts to perturb
the members of this family to exact solutions. There is a
parameter $\eta$ which measures the size of the neck, or gluing
region; the main difficulty is caused by the tension between the
competing demands that the approximate solutions become more
nearly exact as $\eta \to 0$ while the underlying geometry and
analysis become more singular. In this approach, the conformal
method of solving the constraints is used, and the solution
involves a conformal factor which is exponentially close to $1$
(as a function of $\eta$) away from the neck region, but which is
nonetheless not completely localised.

Consider first an asymptotically flat time-symmetric initial data
set, to which several other time-symmetric initial data sets
 have been glued by this method.
If one assumes that the resulting necks are mean outer convex, as
described in detail in Section~\ref{Sah} below\footnote{This will
hold if the gluing regions are made small enough.}, then the
existence of a non-trivial minimal surface, hence of an apparent
horizon, follows by standard results, {\em cf.}\/
Section~\ref{Sah}. This implies the existence of a (possibly
disconnected) black hole region in the maximal globally hyperbolic
development of the data. As for the Brill-Lindquist construction,
the asymptotically flat initial data produced in this way may not
have sufficient differentiability at the resulting $\scrip$ to
obtain good information about $\mcEh$.

Consider, next, hyperboloidal initial data with \bel{kg}
K_{ij}=g_{ij}\;.\ee It follows from \eq{tp} that in this setting
trapped or marginally trapped surfaces are characterised by the
condition \bel{tp1} \theta_+ = H +2\le 0\;.\ee Fixing a polar
coordinate $r$ on the standard three-dimensional hyperboloid, the
constant curvature $-1$ metric takes the form
$$g=\frac 1 {r^2+1} dr^2 + r^2 d\Omega^2$$
(where $d\Omega^2$ is the constant curvature $+1$ metric on
$S^2$), then it is straightforward to calculate that the mean
curvature with respect to the outer normal of the `constant $r$'
geodesic spheres is given by the formula
$$H=2\sqrt{1+r^{-2}}\;.$$
Now suppose we glue together two hyperboloidal initial data sets.
From the point of view of far away observers sitting on the other
side of the ensuing neck, the inner pointing normal for a geodesic
sphere on one half of this configuration is actually pointing
towards them, thus outer-pointing as far as they are concerned;
hence the quantity
$$-H +2=-2/(r^2+\sqrt{r^2+1})$$
measures``trapedness" with respect to the other asymptotic region.
This is negative for any $r>0$ on the hyperboloid, and will remain
strictly negative when $r$ is large enough even after the gluing
has been performed. This means that large spheres on one side of
the neck are trapped from the point of view of the $\scrip$ on the
other side and hence, by standard Lorentzian geometry arguments,
can not be seen from that Scri. This again implies the existence
of a black hole region. Notice that this simplest case is
rotationally invariant, and there is a unique minimal sphere
encircling the neck (see Lemma~\ref{LT1} below). Hence by
continuity, there is at least one marginally trapped, i.e.\ with
$\theta_+ = 0$, rotationally symmetric geodesic sphere. A similar
argument establishes existence of black hole regions when several
initial data sets satisfying \eq{kg}, at least two of which are
hyperboloidal, are glued together.

\section{Outermost apparent horizons with many components}\label{Sah}

In this section we generalise the examples above and consider a
family of asymptotically Euclidean metrics $\{g_\eta\}$, $0 < \eta
\leq \eta_0$, which satisfy the two properties below. Our goal is
to show that when $\eta_0$ is sufficiently small, the outermost
apparent horizon of each of these metrics has a large number of
components.

We assume that for $\eta \in (0,\eta_0]$, $(\hyp,g_\eta)$ is a
Riemannian manifold with boundary of dimension $3$ with a single
asymptotically Euclidean end $E$, and such that $\del \hyp$ is a
union of $I$ copies of $S^2$. (In the context of the initial data
of Section~\ref{Sid}, this amounts to removing from the manifold
that part which lies on the other side of the connecting necks.)
We suppose furthermore that around each boundary component there
is an annular `neck region' $A_i$, $i = 1, \ldots, I$, equipped
with a diffeomorphism $\Phi_i:S^2 \times [-1,1] \to A_i$, such
that $\Phi_i(S^2 \times \{-1\}) = (\del \hyp) \cap A_i$. Thus
$$
\hyp = E(\eta) \cup A_1 \cup \ldots \cup A_I
$$
is a union of manifolds with boundary, intersecting only along the
submanifolds $(\del E(\eta)) \cap A_i= \Phi_i(S^2 \times \{+1\})$
so that the $A_i$ are mutually disjoint. We call $\Phi_i(S^2
\times \{-1\})$ and $\Phi_i(S^2 \times \{+1\})$ the outer and
inner boundaries of $A_i$. The end $E(\eta)$ is diffeomorphic to
an exterior region in $\R^3$, and we fix a family of
diffeomorphisms
\[
\Psi_\eta: \R^3 \setminus \cup_i B(\vec x_i,\rho_i(\eta))
\longrightarrow E(\eta),
\]
and assume that the radii of these balls $\rho_i(\eta)$ tend to
$0$ as $\eta \searrow 0$. (It is only a matter of convention that
we think of the annular regions as fixed, whereas $E(\eta)$ is
identified with an $\eta$-dependent region. However, the metric
$g_\eta$ varies nontrivially on each of these regions.)

Our hypotheses on the metrics $g_\eta$ are as follows:

\begin{itemize}
\item[a)] {\bf[Metric convergence on the distinguished end:]} If
$K$ is any compact subset of $\R^3\setminus \cup_i\{\vec
x_i\}_{i=1,\cdots,I}$, then for some $\alpha\in(0,1)$
$$
\lim_{\eta \to 0} \|\Psi_\eta^*(g_\eta) -
\delta\|_{C^{2,\alpha}(K)} =0\,;
$$
here $\delta$ is the Euclidean metric on $\R^3$. \item[b)] {\bf
[Mean outer convex necks and small minimising cycles:]} For $\eta$
in a sufficiently small interval $(0,\eta_0)$, both the inner and
outer boundaries $\Phi_i^{-1}(S^2 \times \{\pm 1\})$ of $A_i$ are
mean outer convex with respect to $g_\eta$; furthermore, there
exists a smoothly embedded sphere $S_i$ which represents the
fundamental class $\sigma_i \in H_2(A_i,{\mathbb Z})$ and with
area $|S_i| \to 0$ as $\eta \to 0$.
\end{itemize}

Each of the three constructions outlined in Section~\ref{Sid}
produce families of metrics satisfying these hypotheses. For
example, for the construction in Section~\ref{Sid}, if $\vec
M_0:=(m_0,m_1,\ldots,m_{2N})$ is a $(2N+1)$-tuple of nonnegative
numbers and $\vec M(\eta)=(m(\eta),\eta \vec M_0)$ is the
associated mass-parameter vector from that construction, then
$g_\eta:= \hat g_{\vec M(\eta)}$ satisfies both these hypotheses.
Similarly, the initial data of Section~\ref{Sbhgm} satisfy the
hypotheses here if $\eta$ is a sufficiently  small parameter
controlling the outer radii of the $I$ necks across which the
gluing is performed.

\medskip

We begin with a geometric result which holds under slightly more
general hypotheses:

\begin{Lemma} Let $g$ be a Riemannian metric on $A = S^2 \times [-1,1]$
such that the two boundaries $S^2 \times \{\pm 1\}$ are mean outer
convex. Fix a generator $\sigma_A$ for $H_2(A,{\mathbb Z})$. Then
any surface $\Sigma$ which is absolutely area minimising in this
homology class is smoothly embedded, lies in the interior of $A$,
and consists of a single component of multiplicity one.
\end{Lemma}

\proof The existence of a homological area-minimiser $\Sigma$ in
the class of integral currents in a manifold with mean outer
convex boundaries, and the regularity of its support, is a
standard result in geometric measure theory, {\em
cf.}\/~\cite[Theorems 37.2 and 37.7]{SimonGMT}. (These arguments
work equally well for domains with mean outer convex boundaries,
{\em cf.}\/~\cite{SchoenYauIncomp}, and by the maximum principle,
the support of the resulting minimiser is disjoint from $\del A$.)
In particular, the support of $\Sigma$ is a finite union of
smooth, oriented, connected surfaces $\Sigma_1, \ldots ,\Sigma_J$,
where each $\Sigma_j$ appears with some non-vanishing integer
multiplicity $k_j$. Thus on the level of homology
\[
k_1 [\Sigma_1] + \ldots + k_J [\Sigma_J] = \sigma_A,
\]
whereas
\begin{equation}
|\Sigma| = |k_1|\,|\Sigma_1| + \ldots + |k_J|\, |\Sigma_J|.
\label{eq:area}
\end{equation}

We claim that the support of $\Sigma$ has only one component, and
this occurs with multiplicity $1$. To prove this, note first that
any component $\Sigma_j$ divides $S^2 \times [-1,1]$ into
precisely two components. This may be seen by `capping off' the
boundary $S^2 \times \{-1\}$ of $A$ by adding a $3$-ball; the
interior of the resulting manifold $A \cup B^3$ is diffeomorphic
to ${\mathbb R}^3$. By the Jordan separation theorem, any smooth,
oriented, connected surface $\Sigma_j$ embedded in $A$, hence in
$\R^3$, divides this space into an `inside' and an `outside'. For
example, a point $p$ lies in the inner component if (all) generic
paths $\gamma$ connecting $p$ to the outer boundary $S^2 \times
\{1\}$ intersect $\Sigma_j$ an odd number of times. In any case,
this decomposition shows that in homology, $[\Sigma_j] = \pm
\sigma_A$ or else $[\Sigma_j] = 0$ for each $j$. If any $\Sigma_j$
is null-homologous, then we can obviously discard it, since it
adds a positive amount to the area of $\Sigma$ without
contributing to the homology class; possibly changing
orientations, we can therefore assume that each $[\Sigma_j] =
\sigma_A$.

Finally, amongst the $\Sigma_j$ select one, $\Sigma'$, with
smallest area. Then from (\ref{eq:area}), $|\Sigma'| \leq
|\Sigma|$, and equality holds only if $\Sigma'$ is the only
component, and occurs with multiplicity $1$. Thus $\Sigma'$ is the
connected homological area-minimiser, as required. \qed

Now let us return to the more general situation. Using this lemma,
we represent the generator $\sigma_j = [\Phi_j(S^2 \times
\{+1\})]$ of $H_2(A_j,{\mathbb Z})$ by a homologically area
minimising surface $\Sigma_j$; according to hypothesis b),
$\sigma_j$ is also represented by the sphere $S_j$. Both
$\Sigma_j$ and $S_j$ are smoothly embedded, connected surfaces of
multiplicity one. (Since $g_\eta$ has nonnegative scalar
curvature, it is known \cite{SchoenYauIncomp,MR2001j:53051} that
$\Sigma_j$ -- or indeed any stable minimal surface --  must be
either a sphere $S^2$, or possibly a torus $T^2$ if $g_\eta$ is
flat in a neighborhood of $\Sigma_j$.) By assumption, $|S_j| \to
0$, and hence $|\Sigma_j| \to 0$ as well.

It is proved in~\cite[Lemma~4.1]{HI2} that with the hypotheses
above, for every $0 < \eta \leq \eta_0$ there exists a unique
outermost minimal surface $S_\eta$, which is a union of embedded
stable minimal spheres of class ${\mathcal C}^{k+1,\alpha}$ if
$g_\eta$ is of class ${\mathcal C}^{k,\alpha}$. Furthermore, if we
denote by $\hyp'$ the exterior of $S_\eta$ in $\hyp$ (i.e. the
unbounded component of $\hyp \setminus S_\eta$), then $S_\eta$ is
absolutely area minimising in its homology class in $\hyp'$ and
moreover, $\hyp'$ is simply connected.

\begin{Theorem}\label{T1} There exists $\eta_1\in (0,\eta_0]$ such
that if $\eta\in (0,\eta_1]$, then $S_\eta \subset \cup_{i=1}^I
A_i$ and the intersection of $S_\eta$ with each annular region
$A_i$ is nonempty. Hence $S_\eta$ has at least $I$ connected
components. If we assume that there do not exist any stable
minimal homologically trivial surfaces in any of the regions
$(A_i,g_\eta)$ when $\eta$ is small enough, then $S_\eta \cap A_i$
contains exactly one component, and hence $S_\eta$ has precisely
$I$ components.
\end{Theorem}

\proof Let $S(0,R)$ denote a large sphere in $\R^3$ which contains
all of the points $\vec x_i$, and let $\Omega$ denote the part of
$\hyp$ interior to this sphere. Coherently orienting the
fundamental classes $\sigma_j(H_2(A_j,{\mathbb Z}))$, we have that
$[S(0,R)] = \sigma := \sigma_1 + \ldots + \sigma_I$, where we
regard $\sigma_j \in H_2(A_j,{\mathbb Z}) \hookrightarrow
H_2(\hyp,{\mathbb Z})$, as induced by the inclusions $A_j
\hookrightarrow \hyp$. From \cite[Lemma~4.1]{HI2}, we know that
$\hyp'$ is diffeomorphic to the complement of a finite number of
spheres in $\R^3$, and hence $S_\eta$ must be homologous to
$S(0,R)$ as well, i.e.\ $[S_\eta]  = \sigma$. For each $\eta$ we
choose area-minimising representatives $\Sigma_j(\eta)$ of
$\sigma_j$ in $A_j$, as in the preceding Lemma. By hypothesis a),
$S(0,R)$ is mean outer convex for $g_\eta$ if $\eta$ is small
enough, since it is strictly convex for the limiting Euclidean
metric $\delta$. Thus we have
\[
|\Sigma_1| + \ldots + |\Sigma_I| \leq |S_\eta| \leq |S(0,R)|.
\]
The first inequality holds because $\cup \Sigma_i$ is absolutely
area minimising in its homology class in $\hyp$, while the second
inequality follows from the fact that $S_\eta$ is absolutely
minimising in its homology class in $\hyp'$. We claim that for
$\eta$ sufficiently small, $S_\eta$ lies in the union $A_1 \cup
\ldots \cup A_I$. Granting this claim for the moment, let us prove
that $S_\eta$ has at least $I$ components. Choose for each $j$ a
smooth embedded curve $\gamma_j$ which connects the inner boundary
$\Phi_j(S^2 \times \{-1\})$ of $A_j$ to $S(0,R)$, does not
intersect any of the other annular regions $A_i$, $i \neq j$, and
which represents the Poincar\'e dual of $\sigma_j$ in $H_1(\Omega,
\partial \Omega)$. Then the homological intersection number of
$\gamma_j$ with $[S_\eta]$ equals
\[
\langle [\gamma_j], \sigma \rangle = \langle [\gamma_j], \sigma_1
+ \ldots + \sigma_I \rangle = 1.
\]
On the other hand, if $\gamma_j$ is in general position, then this
intersection number is also computed by counting the signed
geometric intersections of this curve and this surface. Therefore
this geometric intersection is nontrivial, which shows that
$S_\eta \cap A_j \neq \emptyset$ for each $j$, and hence $S_\eta$
has at least $I$ components.

To prove the claim, suppose there exists a sequence $\eta_\ell \to
0$ such that $S(\ell) := S_{\eta_\ell}$ contains a point $\vec
q_\ell \in \Omega \setminus \cup A_i$ with $\vec q_\ell \to \vec q
\in \R^3 \setminus \{\vec x_1, \ldots , \vec x_I\}$. The interior
curvature estimate for embedded stable minimal surfaces proved by
Schoen~\cite{Sch}  states that there is a uniform upper bound for
norm squared of the second fundamental of $S(\ell)$ with respect
to $g_\eta$ near $q_\ell$. More precisely, for any $\vec p \in
S(\ell)$ with $\rho(\vec p) = \min_i\{|\vec p - \rho_i(\eta)|\}
\geq \delta > 0$ for $\ell$ sufficiently large, there exists a
constant $C > 0$, independent of $\ell$, such that
$|II_{S(\ell)}(p)|^2 \leq C$. By standard calculus, this implies
that the portion of $S(\ell)$ in a ball of radius $\rho(\vec p/2)$
around $\vec p$ may be written as a graph with uniformly bounded
gradient over a disk of radius $\rho(\vec p)/4$ in $T_{\vec
p}S(\ell)$. In particular, the area of $S(\ell)$ is uniformly
bounded below by a positive constant.

Applying these bounds to a finite covering of $\Omega \setminus
\cup_i B(\vec x_i,\rho)$ for any $\rho > 0$, and then taking a
diagonal subsequence for some sequence $\rho_j \to 0$, we may
extract a subsequence $S(\ell')$ which converges to a {\it
nontrivial} smoothly embedded minimal surface $S(\infty)$ in $\R^3
\setminus \{\vec x_1, \ldots, \vec x_I\}$. Since all of the
$S(\ell')$ are unions of spheres, and the number of components is
uniformly bounded, the limiting surface must have finite genus. In
addition, $S(\infty)$ is compact and has bounded area. We may now
apply a well-known removable singularities theorem for minimal
surfaces, see \cite[Prop.~1]{Choi-Schoen} for a proof, which shows
that $S(\infty)$ is a nontrivial compact embedded minimal surface
in $\R^3$. Since no such surfaces exist, we have reached a
contradiction. We have now proved the first assertion, and hence
that $S_\eta$ has at least one connected component in each $A_i$.

For the remaining assertion, write $S_i(\eta) = S_\eta \cap A_i$,
and suppose that this surface has more than one component for some
$i$, i.e.\ $S_i(\eta) = \cup_{j=1}^J S_{ij}(\eta)$, where $J > 1$
and the $S_{ij}(\eta)$ are smooth embedded surfaces. By the same
argument as in Lemma 3.1, each $S_{ij}(\eta)$ separates $A_i$ into
two components. If $A_i$ contains no null-homologous stable
minimal surfaces, then each component of $A_i \setminus
S_{ij}(\eta)$ must contain exactly one of the two boundaries
$\Phi_i(S^2 \times \{\pm 1\})$. However, the components
$S_{ij}(\eta)$ are disjoint, and so if there are at least two,
then any one must be contained in either the interior or exterior
region of another; since their union is an outermost surface this
is impossible. We conclude that $S_i(\eta)$ is connected. This
completes the proof. \qed

In the case of data of Section~\ref{Sid} the hypotheses of the
second part of Theorem~\ref{T1} are verified:

\begin{corollary}\label{CT1} Let $I\in \N$, $\vec M_0 \in \R^I$, and consider
initial data of Section~\ref{Sid}, with $\vec
M(\eta)=(m(\eta),\eta \vec M_0)$ and $g_\eta:= \hat g_{\vec
M(\eta)}$. If $\eta$ is small enough, than the outermost apparent
horizon is precisely the union of the Schwarzschild horizons
$|\vec x-\vec x_i|=m_i/2$.
\end{corollary}

\proof Let $A_i$ be small annular regions around the $\vec x_i$'s,
chosen so that the metric is exactly Schwarzschild there, then by
Theorem~\ref{T1} we have $S_\eta\subset\cup_iA_i$ for $\eta$ small
enough. The result follows now from the following fact:\qed

\begin{Lemma}\label{LT1}
The only compact embedded minimal surface in a Riemannian
Schwarzschild metric \eq{1} is the sphere $|\vec x-\vec x_0|=m/2$.
\end{Lemma}

\proof The Riemannian Schwarzschild metric is foliated by spheres
of constant mean curvature. These are outer mean convex with
respect to the normal pointing away from the neck. We may now
apply the maximum principle. If $S$ is any compact embedded (or
even immersed) minimal surface, then there is some outermost such
sphere which makes `first contact' with $S$, which is a
contradiction. The only alternative is that $S$ coincides with one
of these spheres, and since it is minimal, it must be the central
one.

We may also argue using Lorentzian methods. In fact, standard
causality theory shows that a compact embedded minimal surface
within a time symmetric Cauchy surface cannot be seen from
$\scrip$, and so we may obtain the conclusion by inspecting the
well known conformal diagram for the Kruskal-Szekeres extension of
the Schwarzschild space-time. \qed

Using \cite[Lemma~4.1]{HI2} one last time, each component of $S_i$
is a sphere, and it is plausible that these must agree with the
homologically area-minimising surfaces $\Sigma_i \subset A_i$,
whose topology is a priori either that of a sphere or a torus. In
each of the examples in the last section, the annular regions
$A_i$ are small perturbations of rescalings of the Riemannian
Schwarzschild metric, and so one may construct a foliation by
constant mean curvature spheres using the implicit function
theorem; from this it follows just as before that there is a
unique stable minimal surface representing $\sigma_i$, so that
$S_i' = \Sigma_i$ for all $i$. However, it is not clear that this
is true in more general cases.

 There is an
analogue of Theorem~\ref{T1} concerning trapped surfaces for
asymptotically hyperboloidal initial data sets. Suppose that
$\hyp$ has the same topology as before, but that the metrics
$g_\eta$ are asymptotically hyperboloidal. Metrics of this sort,
with many necks, can be constructed as in Section~\ref{Sbhgm}. We
suppose that the diffeomorphism $\Psi_\eta^{-1}$ identifies
$E(\eta)$ with the complement of a finite number of balls in
${\mathbb H}^3$ (or indeed any asymptotically hyperboloidal
manifold with constant negative scalar curvature); we also replace
the hypotheses a) and b) by:
\begin{itemize}
\item[a')] {\bf [Metric convergence on the distinguished end:]} If
$K$ is any compact subset of ${\mathbb H}^3\setminus \cup_i\{\vec
x_i\}_{i=1,\cdots,I}$, then for some $\alpha\in(0,1)$
$$
\lim_{\eta \to 0} \|\Psi_\eta^*(g_\eta) -
\ghyp\|_{C^{2,\alpha}(K)} =0\,;
$$
here $\ghyp$ is the standard hyperbolic metric on ${\mathbb H}^3$.
\item[b')] {\bf [Neck boundaries with controlled mean curvature:]}
For $\eta$ in a sufficiently small interval $(0,\eta_0)$, the
outer boundaries $\Phi_i(S^2 \times \{-1\})$ have mean curvature
$h < -2$ (with respect to the inward-pointing unit normal).
\end{itemize}

We shall be using the maximum principle in the following form. Let
$S_1$ and $S_2$ be two oriented, connected, embedded surfaces with
constant mean curvature $H_1$ and $H_2$, respectively. Suppose
that these surfaces are tangent at a point $p$ and their normals
are equal at this point, and that in some small neighborhood $S_1$
lies on the `interior' of $S_2$ (with respect to the normal). Then
necessarily $H_1 \geq H_2$, and if $H_1 = H_2$, these surfaces
must coincide. As a slightly weaker statement, if $H_1$ and $H_2$
are now possibly variable and if $H_1 > H_2$ everywhere, then this
one-sided tangency cannot occur. As an immediate application, let
$\Sigma$ be any compact oriented surface in ${\mathbb H}^3$ which
contains all of the points $\vec x_i$ in its interior, and which
has mean curvature everywhere greater than $-2$ with respect to
its outward normal. (For example, we could let $\Sigma = S(0,R)$,
a large sphere.) This mean curvature remains greater than $-2$
when computed with respect to the metric $g_\eta$ when $\eta$ is
small enough. Hence $S_\eta$ cannot be internally tangent to this
sphere, and this shows that in particular $S_\eta$ is contained in
a fixed neighborhood of the convex hull of the $\vec x_i$.

\begin{Proposition}\label{Pr1h} Under hypotheses a') and b'), there is
at least one (smooth, embedded, oriented) surface $S_\eta$ which
is homologous to $S(0,R) \subset {\mathbb H}^3$ (for sufficiently
large $R$) and which has mean curvature $-2$ with respect to the
normal pointing into the unbounded component of $\hyp \setminus
S_\eta$, i.e. is marginally trapped.
\end{Proposition}

\proof Since $\hyp$ is a manifold with boundary, the volume form
$dV_{g_\eta}$ is exact, hence equals $d\Lambda$ for some
(non-unique) $2$-form $\Lambda$. Now define the functional
\[
L(S) = A(S) + \int_S \Lambda,
\]
Note that changing $\Lambda$ alters $L$ by a constant in each
homology class, but this is irrelevant for our purposes. This
functional was studied, for example, in \cite{WittenYau}, and it
follows from (2.14) in that paper that if $S$ is a smooth
stationary point of $L$, then the mean curvature of $S$ is equal
to $-2$.

Henceforth, let $S(0,R)$ denote any large geodesic sphere in
${\mathbb H}^3$ which encloses all of the points $\vec x_i$, and
which we identify with a surface in $\hyp$ using $\Psi_\eta$. We
may apply the usual geometric measure theory arguments, as
follows, to conclude the existence of a smooth minimiser in the
homology class of $S(0,R)$. First, it is clear that $L(S(0,R))$
increases without bound as $R \to \infty$. Next, when looking for
a minimiser $S$, we may as well assume that $S$ lies in the
bounded component $U$ of $\hyp \setminus S(0,R)$, for if this were
not the case, we could replace $S$ by a homologous surface $S'$ on
which $L$ assumes a smaller value. For example, if $V$ is the
bounded component of $\hyp \setminus S$, then $\partial
(\overline{U \cap V})$ is a suitable\footnote{This follows from
convexity: if one lets $S_1$ be the portion of $S$ outside the
sphere, and $\Pi$ the projection from the exterior onto the
surface of the sphere, then $\Pi(S_1)$ has less area than $S_1$,
because the Jacobian of $\Pi$ is everywhere less than $1$. So the
sphere contribution to $L$ is reduced; clearly the volume
contribution is reduced as well.} choice for $S'$.  Hence, since
we may assume that any minimising sequence $S_j$ remains within a
compact set in $\hyp$, and since $L$ is bounded below, we may find
a minimiser $S_\eta$. The assumption that the outer boundaries
have mean curvature $H < -2$ ensures that $S_\eta$ remains in the
interior of $\hyp$, {\em cf.}\/~\cite[Lemma 4]{WittenYau}. The
same regularity theory as was quoted earlier implies that the
minimiser $S_\eta$ is a smooth embedded and oriented surface in
the interior of $\hyp$. \qed

\begin{Theorem}\label{T1h} Assume $(\hyp,g_\eta)$ is asymptotically
hyperboloidal and satisfies the hypotheses a') and b'). For $\eta$
in some sufficiently small interval $(0,\eta_1]$, any trapped
surface $S_\eta$ which is homologous to $S(0,R)$ is contained in
$\cup_{i=1}^I A_i$ and has at least $I$ connected components.
\end{Theorem}

Notice that we are not assuming that $S_\eta$ is an outermost
trapped surface here.

\proof We have already indicated that such trapped surfaces exist.
To prove that $S_\eta \subset \cup_i A_i$, we proceed as before
and assume that this is not the case. To take a limit as $\eta \to
0$, we use the methods and estimates from~\cite{KKMS}
and~\cite{KK}, which adapt in a straightforward way to small
metric perturbations of hyperbolic space, {\em cf.}\/
also~\cite[Lemma~2]{WittenYau}. In general, the situation is not
as simple as for stable minimal surfaces because of the
possibility of small necks in $S_\eta$ pinching off, even in
regions where the ambient geometry is uniform. One can prove that
the limit surface $S'$ is a finite union of smooth embedded
surfaces $S'_j$ which are mutually tangent at their points of
intersection. (This part of the argument does not use specifically
that $|H|=2$, and it is possible one could use this special
feature more strongly and show directly that $S'$ is smooth;
however, this is not so important for our purposes.) We may use
the same removable singularities theorem as before, or rather its
proof, to show that each of the $S_j'$ are smooth at the points
$\vec x_i$. However, each $S'_j$ is compact and has constant mean
curvature $-2$. But one could then find a horosphere tangent to
$S_j'$, for example by bringing it in from infinity (in any
direction) until it reaches a point of first contact, and this
would contradict the maximum principle. Hence $S_j'$ could not
exist. (An alternative nonexistence proof is to note that if such
$S_j$'s existed, then Minkowski space-time would contain non-empty
black hole regions.)

We have now reduced to the case where $S_\eta \subset \cup A_i$.
The same intersection theory argument as in the proof of
Theorem~3.2 shows that each of the intersections $S_\eta \cap A_i$
is nonempty, and so $S_\eta$ must have at least $I$ components.
Note that each $A_i$ contains an area-minimising surface
$\Sigma_i$ which is homologous to the outer boundary, and the
maximum principle implies that $S_\eta$ is contained in the region
between $\Phi_i(S^2 \times \{-1\}$ and $\Sigma_i$. \qed

One can impose various geometric conditions on the metric $g_\eta$
on the $A_i$ which would ensure that $S_\eta$ has exactly $I$
components. A rather stringent one, which however is satisfied for
the asymptotically hyperboloidal initial data sets of \cite{IMP}
for $\delta$ small enough, is:
\begin{itemize}
\item[c')] The diffeomorphisms $\Phi_i$ can be chosen, now
possibly depending on $\eta$, so that each sphere $\Phi_i(S^2
\times \{t\})$ has constant mean curvature $H_i(t)$, and that each
$H_i$ is a monotone function on $[-1,1]$ with values in some
interval $[-h(\eta),h(\eta)]$, where $h(\eta) > 2$.
\end{itemize}

To see that the initial data sets of \S 2.3 have CMC foliations on
each neck region, one can argue as follows. The quantitative
estimates for the metric $g_\eta$ on these neck regions from
\cite[\S 8]{IMP} show that if we scale $(A_i,g_\eta)$ to have a
fixed neck size (e.g.\ to have injectivity radius always equal to
$1$), then this annulus is ${\mathcal C}^2$ quasi-isometric, with
constant tending quickly to $1$ as $\eta \to 0$, with the neck
region for the Riemannian Schwarzschild space (scaled to have the
same normalisation). This latter space has a global CMC foliation,
and by the implicit function theorem we can produce such a CMC
foliation in any fixed neighborhood of the neck. The outermost
leaves of this foliation will have mean curvature $\pm h$, say,
and when rescaled down to the original size, these leaves now have
mean curvature $\pm h(\eta)$, where $h(\eta) \to \infty$.

We use this CMC foliation as follows. Consider the component
$S_{i,\eta}  = S_\eta \cap A_i$. Choose $\tau'$ and $\tau''$ so
that $S_{i,\eta} \subset S^2 \times [\tau',\tau'']$, and such that
this is the narrowest band with this property. Then $S_{i,\eta}$
is tangent to both boundaries, and its outward unit normal at
these points lies in the same direction as $\partial_t$. Denoting
by $H'$ and $H''$ the constant mean curvatures of those two
boundaries, then the maximum principle gives that $H' \geq -2 \geq
H''$. But $\tau' \leq \tau''$ and so $H' = H''$ and finally
$S_{i,\eta}$ must coincide with a leaf of the foliation, and hence
is connected.

\section{Sections of event horizons have at least $I$ components}
\label{Seh}

In this section we analyze the global structure of the maximal
globally hyperbolic developments of families of  initial data
sharing certain overall properties with those of
Section~\ref{Sid}, when the mass parameters are sufficiently
small. This question is rather different from the one raised in
the previous section, because the existence of apparent horizons
involves only the geometry of the initial data, which is fairly
well controlled. On the other hand, the notion of the event
horizon involves the global structure of the resulting space-time,
about which only very scant information is available. Before
proceeding further, the following should be said: because gravity
is attractive, and because the Schwarzschild regions of the
initial data of Section~\ref{Sid} are initially at rest with
respect to each other, one expects that those regions will ``start
moving towards each other", leading either to the formation of
naked singularities, or to a single black hole. In particular the
resulting event horizon, if occurring, is expected to be a
connected hypersurface in space-time. Nevertheless, the properties
of the maximal globally hyperbolic developments $(\mcM,g)$ of the
data which we present below lead us to conjecture that \emph{there
exists no slicing of $\mcM$ by Cauchy surfaces $\hyp_\tau$ which
are asymptotically flat in all their asymptotic regions and in
which all the intersections $\mcE^+\cap \hyp_\tau$ are connected}.
This seems to be the proper way of making precise the many-black
hole character of certain families of black hole space-times.
While we do not prove such a conjecture, it follows from what is
said below that for some configurations there exist natural
slicings of $\mcM$ which do have this property.
%

\begin{figure}[t]
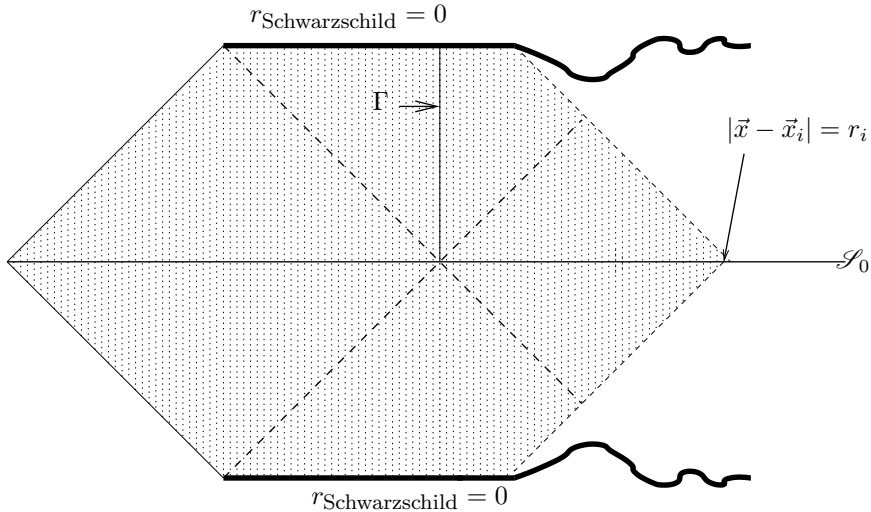

\label{Sfig0}%
\begin{center}
  \include{picture2}
\end{center}
\caption{The maximal globally hyperbolic development of the
initial data of Section~\protect\ref{Sid} in a neighborhood of
$B(\vec x_i,r_i)$. The metric in the dotted
 region is \emph{exactly} the
Schwarzschild metric. The $45$-degrees sloped dashed lines
correspond to the Schwarzschild event horizons. The straight part
of the boldface lines is \emph{exactly} the Schwarzschild
singularity; one expects that some form of that singularity will
survive in  nearby regions influenced by the non-Schwarzschildian
initial data, as depicted by the curved part of the boldface line,
but no results of this kind are known.}
\end{figure}

Recall that the black hole event horizon $\mcE^+$ is usually
defined as
\begin{equation}\label{4.1} \mcE^+:= \pJ ^-(\scrip; (\bmcM,\wtg))\;.
\end{equation}
Here the causal past $J^-$ is taken with respect to the
conformally rescaled space-time metric $\wtg$ on the completed
space-time with boundary $\bmcM:=\mcM\cup\scrip$. Thus, the
starting point of any black hole considerations is the existence
of a conformal completion at future null infinity $\scrip$. In
this context one usually assumes that $\scrip$ satisfies various
completeness conditions~\cite{GerochHorowitz,HE,Waldbook} (compare
the discussion in \cite{ChDGH,Chrusciel:2002mi}). As already
mentioned, for the metrics of Section~\ref{Sid} past-completeness
of $\scrip$ is guaranteed by the fact that the initial data are
exactly Schwarzschild outside of a compact set. However, the
current understanding of the global properties of solutions of the
Cauchy problem for the Einstein equations is insufficient to
guarantee any future completeness properties of the resulting
$\scrip$. Nevertheless, we shall see that for some of those
metrics the conformal boundary $\scrip$ can be chosen sufficiently
large to the future so that $\mcEh$ defined by \eq{0} will have
more than one component. (This feature will persist upon enlarging
$\scrip$, and will therefore also hold for a maximal one.) Before
passing to a proof of this fact let us point out that the
existence time of the solution, defined as the lowest upper bound
on the existence time of all geodesics normal to $\hyp$, goes to
zero as the mass parameters go to zero. In order to see that, let
$\Gamma$ be a maximally extended future directed timelike geodesic
normal to $\hyp$ starting at the minimal neck of
 the Einstein-Rosen bridge of the usual Kruskal--Szekeres
 extension $\Sdev$ of the Schwarzschild space-time with mass $m$. Either an explicit calculation,
 or a simple
 scaling argument, show that the Lorentzian length of $\Gamma$ is
 proportional to $m$. Now, if $\delta$ in \eq{mcondi} is small
 enough, then the maximal globally hyperbolic development
 $(\mcM,g)$ of the initial data of Section~\ref{Sid} will contain
 a region isometrically diffeomorphic to a neighborhood of
 $\Gamma$ in $\Sdev$ as in Figure~\ref{Sfig0}.
 This shows that for small data $(\mcM,g)$ is necessarily
 ``small", in the sense made precise above, and a complete understanding of the global structure of
 the resulting space-times might be
 a delicate issue.

Let us return to the problem of main interest here, namely
non-connectedness of sections of $\mcEh$, as defined by \eq{0}. We
shall show  that the stability results of
Friedrich~\cite{Friedrich} can be used to reduce this  question
 to elementary considerations of light-cones in
Minkowski space-time. Recall that the simplest conformal
completion of the timelike future of a point in Minkowski
space-time  $\Mink$ is obtained by performing the space-time
inversion \bel{4.2} \{x^0>0\;,\ \eta_{\alpha\beta}x^\alpha x^\beta
<0\}\ni x^\mu \to y^\mu = \frac {x^\mu}{\eta_{\alpha\beta}x^\alpha
x^\beta}\in \{y^0<0\;,\ \eta_{\alpha\beta}y^\alpha y^\beta
<0\}\;.\ee Here $\eta_{\mu\nu}$ is the Minkowski metric. A
drawback of the transformation \eq{4.2} is that it does not give
the whole conformal completion of Minkowski space-time at once;
however, a major advantage thereof is that the rescaled metric is
again the Minkowski one, so that the causal properties of the
rescaled space-time are straightforward to analyse, and to
visualise:$$\eta_{\mu\nu}dx^\mu dx^\nu= \frac 1
{(\eta_{\alpha\beta} y^\alpha y^\beta)^2}\eta_{\mu\nu}dy^\mu
dy^\nu\;.
$$ Under \eq{4.2} the future timelike cone $I^+(0_x;\Mink)$ of the
origin $0_x$ of the $x^\mu$ coordinates becomes the past timelike
cone $I^-(0_y;\Mink)$ of the origin $0_y$ of the $y^\alpha$
coordinates; further, $0_y$ is the future timelike infinity point
$i^+$, while $\pJ ^-(0_y;\Mink)$ becomes that part of the
Minkowskian $\scrip$ which lies to the causal future of $0_x$ in
the conformally completed Minkowski space-time.

\begin{figure}[ht]
\begin{center}
  \includegraphics[scale=2.5,width=\textwidth]{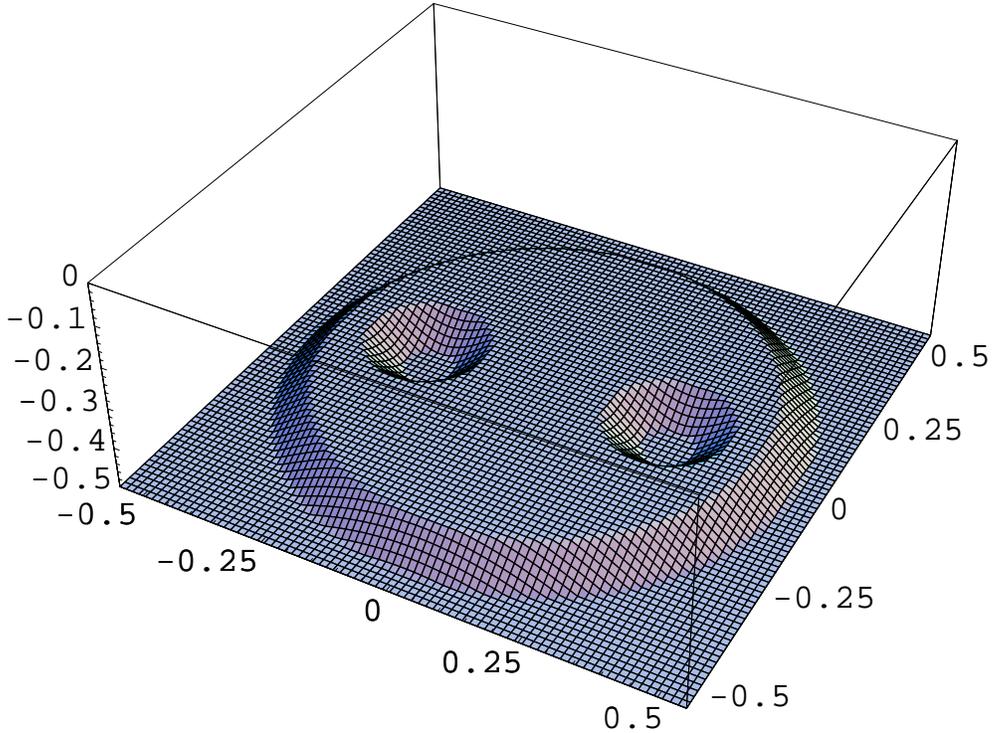}
\caption{A $(2+1)$--dimensional version of the space-time
$\mcM_\tau$, $I=2$, for $\tau$ smaller than the time $\tau_-$ of
\protect\eq{et}.}\label{F41}
\end{center}
\end{figure}

\begin{figure}[ht]
\begin{center}
  \includegraphics[scale=2.5,width=\textwidth]{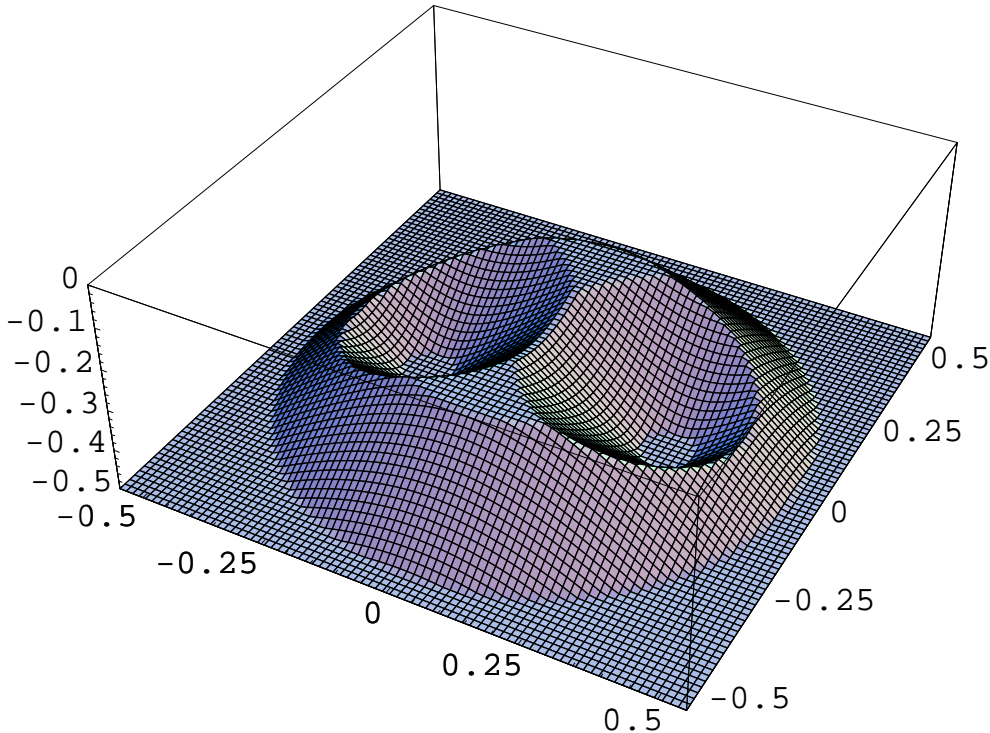}
\caption{The space-time $\mcM_\tau$ of Figure~\protect\ref{F41}
for $\tau_*<\tau<\tau_+$, with $\tau_*$ given by
(\protect\ref{et2}); compare
Figure~\protect\ref{Sfig}.}\label{F42}
\end{center}
\end{figure}

\begin{figure}[t]
\begin{center}
  \includegraphics[scale=2.5,width=\textwidth]{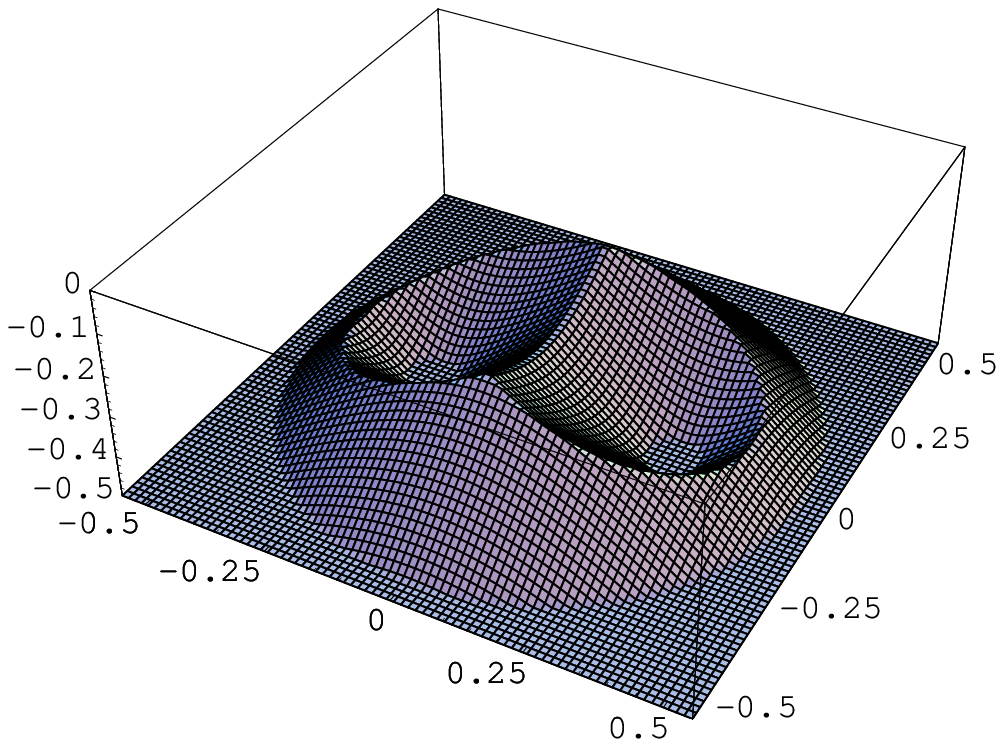}
\caption{The space-time $\mcM_\tau$ of Figure~\protect\ref{F41}
for $\tau$ larger than the time $\tau_+$ of
(\protect\ref{tauplus}).}\label{F21}
\end{center}
\end{figure}

Choose, now, a set of points $\vec y_i$ and strictly positive
numbers $\delta_i$, $i=1,\ldots I$, with \bel{ddef}|\vec
y_i|+\delta_i< 1/2\;,\ \delta_i < |\vec y_i|\;, \ee with the balls
$B(\vec y_i, \delta_i)$ --- pairwise disjoint. The points
 $\vec y_i$ should be thought as the $y$--coordinates equivalents of the points
$\vec x_i$ of Section~\ref{Sid}. Let the initial surface $\hypo$
be defined by the equation\footnote{The value $-1/2$ for $y^0$ is
chosen for definiteness; any other value can of course be chosen.
It is, nevertheless, worthwhile mentioning  that this choice
corresponds to an upper hyperboloid $\{x^0=-1+\sqrt{1+r^2}\}$. It
appears that initial data similar to those of Section~\ref{Sid}
can be constructed directly on such hyperboloids by  extending the
techniques of Corvino and Schoen to a hyperboloidal setting.} \bea
& \hypo = \{y^0=-\frac 12\;,\ 0\le |\vec y| < \frac 12\;, \ \vec y
\not \in B(\vec y_i, \delta_i)\}
 & \eeal{4.3} with $(y^\mu)=(y^0,\vec y)$. Set \bea\nonumber & K_i :=
\{y^0=-\frac 12\;,\ \vec y \in B(\vec y_i, \delta_i)\}\;,&
\\ & \mcM_\tau:=\{-\frac 12 \le y^0\le \tau\}\cap I^-(0_y;\Mink)\setminus \Big(\cup_{i=1}^I J^+(K_i;\Mink)\Big)
\;,&\nonumber\\ 
& \bmcM_\tau:=\{-\frac 12 \le y^0\le \tau\}\cap
J^-(0_y;\Mink)\setminus \Big(\cup_{i=1}^I J^+(K_i;\Mink)\Big)\;.&\nonumber\\
&&\eeal{4.4} (See Figures~\ref{F41}-\ref{F21}.)

The parameter $\tau$ should be thought of as the $y^0$--coordinate
height of the regions on which the solution associated to the
non-trivial initial data exists.  One can think of the regions
$J^+(K_i;\Mink)$ as the regions where the non-trivial geometry,
associated to neighborhoods of the black hole regions, is
localised.

 In order to be able to take advantage of Friedrich's stability
 results~\cite{Friedrich}, we make the following hypotheses: we
 consider families of hyperboloidal initial data sets
$\{(g_\eta,K_\eta)\}_{\eta \in [0,\eta_0]}$, defined on $\iniB
\setminus \cup_i\{\vec y_i\}_{i=1,\cdots,I}$, for some
$\eta_0\in\R^+$, $I\in\N$, such that
\begin{itemize}
\item[a)] {\bf[Uniform convergence]} For any open set $\mcK$ which
has compact closure in $\iniB \setminus \cup_i\{\vec
y_i\}_{i=1,\cdots,I}$ we have\footnote{The $C^k$ norms here can be
replaced by any Sobolev norms which guarantee that the resulting
space-time metric, obtained by evolving the initial data using the
vacuum Einstein equations, is $C^2$.}
$$\lim_{\eta \to 0} \Big(\|g_\eta - \ghyp\|_{C^5(\mcK)} +\|K_\eta-\ghyp \|_{C^4(\mcK)}\Big)=0\;.$$
Here $\ghyp$ stands for the unit hyperbolic metric.

\item[b)] {\bf [Existence of $I$ trapped surfaces]} There exist
$r_i$, $i=1,\ldots,I$ such that for every $\eta$ there exists a
compact smooth embedded trapped or marginally trapped  surface
$S_{i,\eta}\ne \emptyset$ satisfying
$$S_{i,\eta} \subset B(\vec y_i,r_i)\;.$$
The balls $B(\vec y_i,r_i)$ will further be required to be
pairwise disjoint. In some of the arguments below b) will need to
be strengthened to:

 \item[b')] Moreover,
$$\limsup_{\eta\to 0}\{|\vec y-\vec y_i|: {\vec y \in S_{i,\eta}}\} = 0\;.$$
(For marginally trapped surfaces this does follow from a), b), and
from what is said at the end of Section~\ref{Sah}.)
\item[c)]{\bf[Existence of $\scrip$]} The resulting family of
space-times admit  conformal completions which are sufficiently
differentiable so that Friedrich's stability theorem (or perhaps
some extension thereof, in the spirit of
\cite{ChLengardprep,Lengard}) applies.
\end{itemize}

It is not immediately clear whether the initial data of
Section~\ref{Sid} are compatible with those hypotheses. There are
a few issues here. Suppose, for instance, that the solution
associated to the initial data of Section~\ref{Sid} remains as
close as desired to the Minkowski  one, when making the mass
parameters small, in a small neighborhood of the spheres $S(\vec
x_i,r_i)$, for a long time, and that the time in question tends to
infinity as the mass parameters $m_i$ go to zero. In such a case
the hypotheses above would obviously hold, whatever the choice of
the hyperboloidal initial surface $\hyp_0$. However, we are not
aware of any argument which would justify that this is the correct
picture, and the discussion around Figure~\ref{Sfig0} suggests
that this might actually be wrong. A simple way of avoiding the
question of the time of existence of the solution near the spheres
$S(\vec x_i,r_i)$'s is to suppose that all the points $\vec x_i$
lie on the surface of some sphere.\footnote{This involves no loss
of generality if $I=2$, or if $I=3$ and the $\vec x_i$'s are not
aligned. However, for $I=3$ the configurations of
Section~\ref{Sid} are actually co-linear.} We can then choose the
hyperboloid so that the $\vec x_i$'s lie on the intersection of
this hyperboloid with the hypersurface $x^0=0$. For such
configurations clearly all the hypotheses above are satisfied.

In this context the following comment is also appropriate: So far
we have assumed that the initial data are prescribed on the
hypersurface $\hypo$ given by \eq{4.3}. The exact choice of
$\hypo$ is clearly irrelevant, and a similar picture would  be
obtained at this stage with any hypersurface $\hypo$ which
asymptotically approaches a hypersurface of constant conformal
time $y^0$. In particular we could choose $\hypo$ to coincide with
the hypersurface $\{x^0\}=\const$ for some large positive constant
for $|\vec y|<R$, and to coincide with the hypersurface
$\{y^0=-1/2\}$ for $|\vec y|$ large enough. On $\hypo$ we can use
the initial data of Section~\ref{Sid} for $|\vec y|<R$, and
appropriate data obtained by time-evolution elsewhere. In such a
case, the nature of the initial data of Section~\ref{Sid} would
guarantee that the induced data on the new hypersurface would have
almost all the properties used in the discussion above: The only
property missing is that the limiting data set, as $\eta$ tends to
zero, would not be the hyperbolic data set $(g=\ghyp, K=\ghyp)$,
but one corresponding to an appropriate hypersurface in Minkowski
space-time. However, our proof of existence of $I$ components of
the initial section of the event horizons relies on the fact that
the radii $\delta_i$ can be made arbitrarily small on a
hypersurface of constant $y^0$--time, and we wouldn't be able to
achieve the desired conclusions on general hypersurfaces.

Finally, consider the initial data of Section~\ref{Sbhgm} on
asymptotically flat hypersurfaces and on hyperboloids; those
initial data can be chosen to satisfy conditions  a), b) and b')
above.  The initial data of~\cite{IMP} on asymptotically flat
hypersurfaces are not known to satisfy condition c). However,
those constructed in~\cite{IMP} on hyperboloids can be chosen to
satisfy that condition: the resulting globally hyperbolic
developments will not have a $\scrip$ which is complete to the
past, but it should be clear from the arguments below that this is
irrelevant for most of the problems discussed here.

 Returning to our model spaces $\mcM_\tau$,  the corresponding model data on
$\hypo$ are exactly those for the Minkowski metric. In the
physical metric the initial data on $\hypo$ will be close to the
Minkowskian ones, the difference being as small as desired when
$\eta$ is made sufficiently small. Under conditions a) and c) as
spelled-out at the beginning of this section, the usual arguments
about continuous dependence of solutions of hyperbolic PDE's upon
initial data over compact sets, as applied to the conformal
Einstein equations of Friedrich~\cite{Friedrich}, show that for
any fixed $\tau$ the physical metric $g$ will exist on $\mcM_\tau$
and will be as close as desired to the Minkowski one on
$\bmcM_\tau$, when the initial data on $\hypo$ are sufficiently
close to the Minkowski ones.  This implies that the causal
structure of the physical space-time on $\bmcM_\tau$ will be
approximated as accurately as desired by that of the Minkowski
space-time on $\bmcM_\tau$, when the initial data are sufficiently
close to the Minkowskian ones on $\hypo$. In particular the
figures presented here will accurately describe the geometry of
null geodesics in the physical space-time.

\begin{figure}[t]
\begin{center}
\input{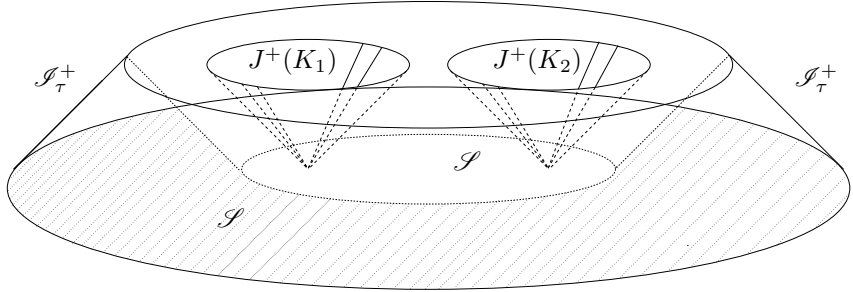}
\caption{$\bmcM_\tau$ for $\tau<\tau_-$, compare
Figure~\protect\ref{F41}. The shaded area is the part  of $\hyp$
which can be seen from $\scrip_\tau$, and its complement in $\hyp$
is therefore the (partly ``physically wrong") black hole region,
within $\hyp$, with respect to $\scrip_\tau$.}\label{Sfig}
\label{F3}
\end{center}
\end{figure}

 Let
\bel{et0}\scrip_\tau:= \{-\frac 12 \le y^0\le \tau\} \cap \pJ
^-(0_y;\Mink)\ee be the conformal boundary of $\mcM_\tau$ and
suppose that \bel{et} \tau < \tau_-:=-\frac 12 +\min_i | \frac 12
- |\vec y_i| - \delta_i|\;.\ee In that case the black hole event
horizon $\mcE_\tau^+$, in the space-time $(\bmcM_\tau,\eta)$,
associated with the conformal boundary $\scrip_\tau$,
$$\mcE_\tau^+:= \pJ ^-(\scrip_\tau; (\bmcM_\tau,\eta))\;,$$
will be a union of spheres:
$$\mcE_\tau^+=\cup_{t\in [-\frac 12,\tau]}\{y^0=t\;,\ |\vec y| =
-2\tau + t\}\;,$$ see Figure~\ref{F3}.
In particular $\mcE_\tau^+$ will be connected, so that each
section thereof through a hypersurface $$\{y^0=\const\}$$ will
also be connected. This holds for the Minkowski metric, and hence
also for the physical metric for initial data sufficiently close
to Minkowskian ones. Thus, if the physical space-time develops a
singularity and stops to exist at some time $\tau$ satisfying
\eq{et}, then the boundary of the black hole region will be
connected. As long as this last possibility occurs it is
meaningless
--- within the $\scrip$ framework --- to assert that $\mcM_\tau$ is a
multi-black-hole space-time.\footnote{On the other hand, at an
intuitive level it is clear that, whatever the value of $\tau$,
the physical space-time does contain distinct regions which
display ``black hole" properties, even though this does not fit
well into the $\scri$ framework. It seems that any significant
insight into such situations will be gained only after better
understanding of the long time behavior of solutions of Einstein
equations will have been reached.} We stress that the global
structure of Figure~\ref{Sfig} could very well arise for
non-trivial initial data, whether small or large, \emph{even if
all singularities are shielded by the event horizon} (in which
case $\tau$, near $\scrip$, can be thought of as being infinite,
and should not be identified with a Minkowskian coordinate). The
point of our considerations below is to show that this will not
happen for some configurations.

Now, as soon as the initial value of $y^0$ exceeds the value
$\tau_-$ given by \eq{et}, some null geodesics starting with this
initial value backwards in time from the Minkowskian Scri
$$\scrip_\Mink:=\pJ ^-(0_y;\Mink)$$ enter the
region $J^+(K_i)$ where the metric fails to be close to the
Minkowski one, even for small mass positive parameters $m_i$, and
where singularities \emph{do} form in short time. The visibility
of those singularities from $\scrip$ would be forbidden if a
suitable version of cosmic censorship hypothesis applied, but  no
such results have been established so far. As of today there is no
justification for the possibility that the physical $\scrip$ can
be continued uniformly  beyond the points at which some of the
generators of the Minkowskian $\scrip$ meet some null geodesics
emanating from the $K_i$'s (though these generators
actually do continue ``a little" in the situation at hand). 
Whatever the case, stability implies that one might continue each
generator of the physical $\scrip$, associated to the non-trivial
initial data, to the future from the boundary of the initial data
hypersurface \emph{up to the first point} the Minkowskian past of
which intersects one of the $K_i$'s. From this point of view the
only significant feature distinguishing  various values of $\tau$
is that sections of the model $\scrip_\tau$, as defined by
\eq{et0}, with hypersurfaces $\{y^0=t\}$ will be spheres for
$t<\tau_-$, \emph{cf.\/}~Figures~\ref{F41} and \ref{Sfig}, while
this will not be the case anymore if
$\tau>t>\tau_-$.\begin{figure}[t]
\begin{center}
\includegraphics[width=0.8\textwidth]{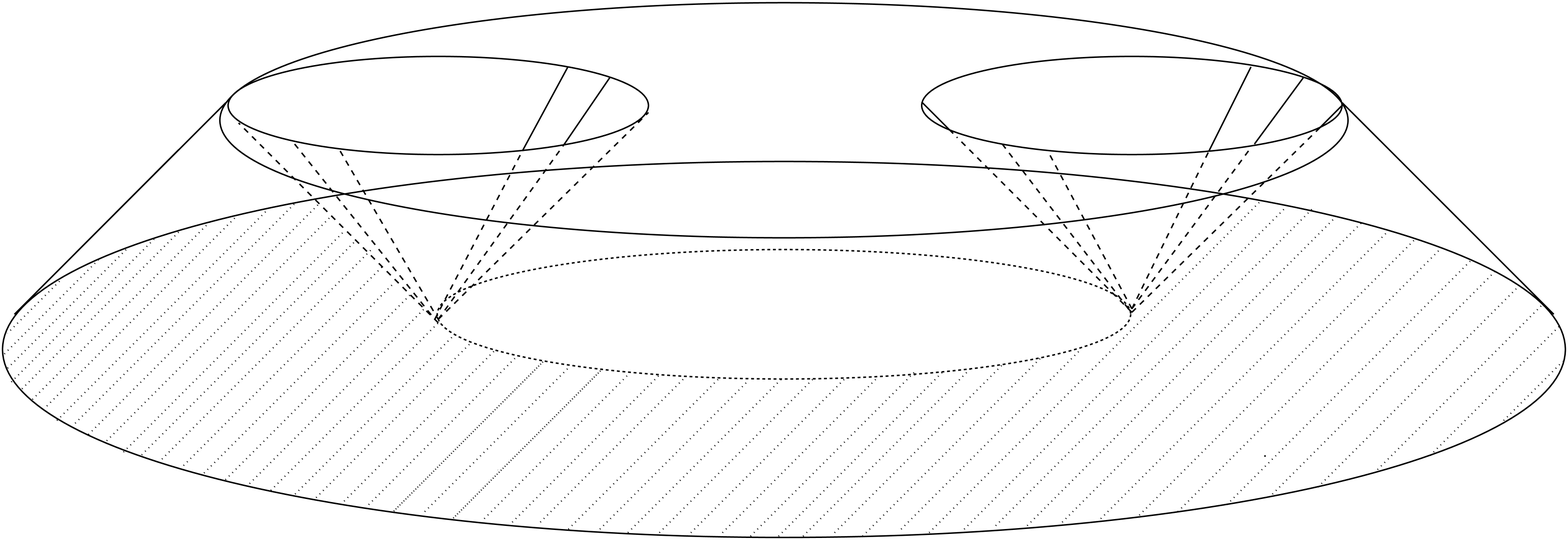}
\end{center}\caption{\protect $\bmcM_\tau$ with $\tau=\tau_-$. The shaded area is the
part  of $\hyp$ which can be seen from $\scrip_\tau$.}\label{F31}
\end{figure}
Furthermore, the causal geometry of $(\bmcM_\tau,\eta)$ becomes
interesting only for \bel{et2} \tau
> \tau_*:= -\frac 12 +\min_i | \frac 12 - |\vec y_i| + \delta_i|\;.\ee
We shall not attempt to analyse exhaustively what happens for all
$\tau>\tau_*$ and all possible  values of $\vec y_i$, but we will
concentrate on a few specific cases. We set
$$\bhz:=\mcE_\tau^+\cap \hyp_0\;.$$
We wish to exhibit configurations for which $\bhz$ has at least
$I$ components. Now, by standard causality theory\footnote{In the
case of the initial data of Section~\ref{Sid} the use of causal
theory is actually not needed: the existence of black hole regions
follows immediately from the Schwarzschildian character of the
data on $B(x_i,r_i)$, as made clear by Figure~\ref{Sfig}.}, the
$S_{i,\eta}$'s of condition b) cannot be seen from $\scrip_\tau$,
whatever the value of $\tau$. It follows that $\bhz$ is never
empty. Our aim is to construct hypersurfaces $\mcN_i\subset
\hyp_0$, $i=1,\ldots,I-1$ with the following properties:
\begin{enumerate}
\item $\mcN_i \subset I^-(\scrip_\tau)$, so that $\bhz\cap
\mcN_i=\emptyset$. \item The $\mcN_i$'s separate $\hyp_0$ into $I$
distinct, open, connected sets $\mcO_i$ such that each $\mcO_i$
contains precisely one $S_{i,\eta}$.
\end{enumerate}
It then clearly follows that $\bhz$ has at least $I$ components.

\begin{figure}[t]
\begin{center}
  \includegraphics[width=\textwidth]{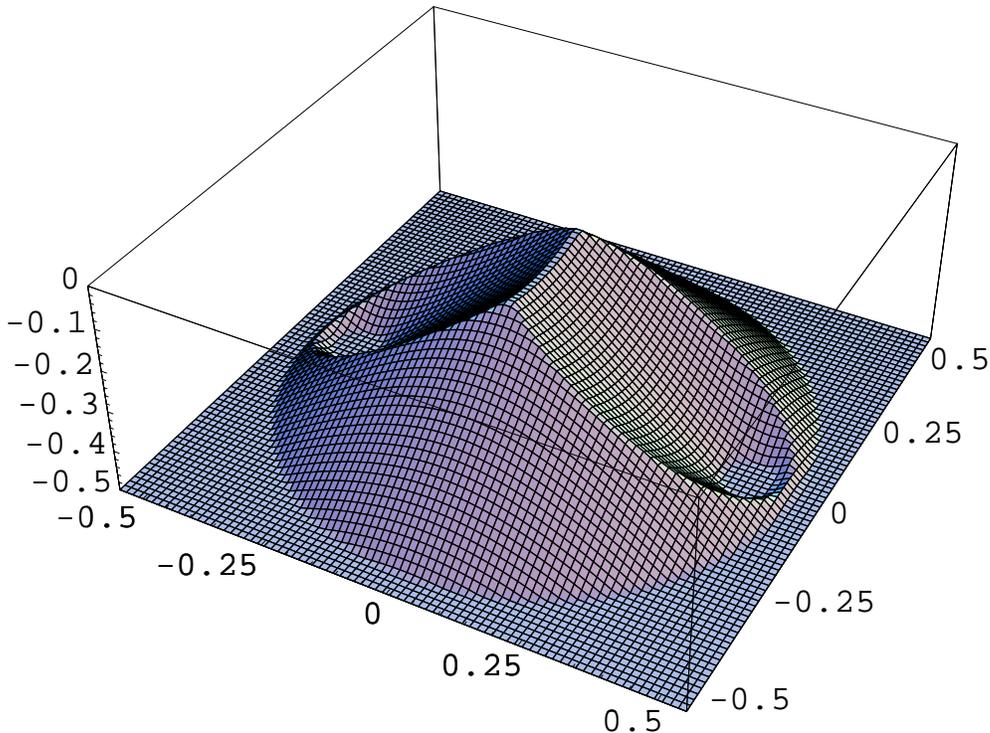}
\caption{\protect $M_0$ for $\vec y_i$'s close to the conformal
boundary.}\label{F25}
\end{center}
\end{figure}

\begin{figure}[h]
  \includegraphics[width=\textwidth]{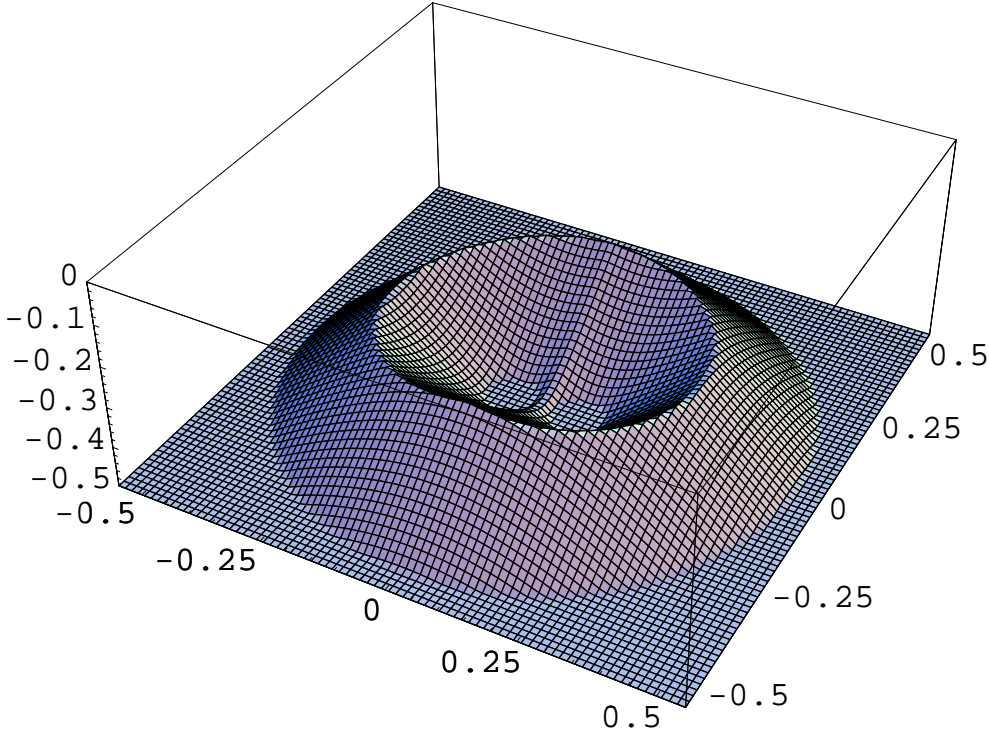}
\caption{$M_0$ for $\vec y_i$'s close to each other.}\label{F23}
\end{figure}

Let us start with the case $I=2$. Without loss of generality one
can then assume $\vec x_1=-\vec x_2$. Further, under the current
hypotheses one can without loss of generality assume that the
constants $\delta_i$ of \eq{ddef} satisfy $\delta_1=\delta_2$ by
replacing the smaller of the $\delta_i$'s by the larger one, and
making the parameter $\eta$ smaller if necessary. From now on we
assume that $\eta$ has been chosen small enough so that the
physical metric exists on $\mcM_\tau$ with $\tau$ larger than
\bel{tauplus} \tau_+:= {a^2-\frac 14}<0\;:\ee this value of $\tau$
corresponds to  the value of $y^0$ at the meeting points of a
generator of $\pJ ^+{(K_1,\Mink)}$ \emph{and} a generator of $\pJ
^+{(K_2,\Mink)}$ \emph{and} a generator of $\pJ ^-{(0_y,\Mink)}$
--- see the proof of Proposition~\ref{Pcriterion} below.
This is also the ``highest point" of $\mcM_\tau$ for $\tau\ge
\tau_+$, compare Figures~\ref{F21}, \ref{F25} and \ref{F23}.
Finally,  this corresponds to the value of $\tau$ above which
$\mcM_\tau$ does not change any more:
$$\forall \tau\ge \tau_+ \quad \mcM_\tau = \mcMs \;,$$
It is then obvious from Figure~\ref{F25}, 
in which the $K_i$'s are very close to the conformal boundary,
that the past of the ``highest points" of $\mcM_\tau$ contains
points lying on the straight line segment connecting the two black
holes.
This same property is still visible, though with a little more
effort, from Figure~\ref{F21} 
where the black hole regions are fairly far away from each other.
On the other hand, it should be clear from
Figure~\ref{F23} 
that $\mcE^+_{\tau_+}(0)$ will be connected there.
Before presenting a precise form of those statements, let us
introduce the notation
$$(y^\mu)=(t,x,y,z)\;.$$
In the proposition that follows the constant $-1/2$ appearing in
\eq{4.3}-\eq{4.4}, representing the $y^0$ coordinate of the
initial data hypersurface, has been replaced by an arbitrary
constant $\tau_0<0$:
 \begin{Proposition}
 \label{Pcriterion} Suppose that  $I=2$, $\delta:=\delta_1=\delta_2$, $\vec
 y_1=-\vec y_2$, set $a:=|\vec y_1|$.
 Then the plane $$\mcN_1:=\{t=\tau_0\;,\; x=0\}\subset \{t=\tau_0\}$$ is included in
 $J^-(\scrips ,(\bmcMs,\eta))$ if and only if
 \bel{ePc1}\frac{2\delta}{|\tau_0|}\le \sqrt{1+\left(
 \frac{2a}{|\tau_0|}\right)^2} - 1\;.\ee
\end{Proposition}
\begin{Remark}\label{R4.2}
\Eq{ePc1} always holds for $\delta/|\tau_0|$ small enough. This is
all that is needed  for our purposes: the sets $K_i$ have been
chosen to contain the non-trivial geometry, and condition a)
guarantees that they can be chosen as small as desired by choosing
the parameter $\eta$ small enough. On the other hand, it \emph{is}
restrictive: for example, for $a/|\tau_0|=1/2$, which is the case
in Figures~\ref{F42}--\ref{F21}, \eq{ePc1} leads to
$\delta\le(\sqrt{2}-1)|\tau_0|/2$, which can fail to be satisfied
without violating our remaining restrictions $\delta<a$,
$\delta+a< |\tau_0|$.
\end{Remark}

\proof Clearly $p_0:=(\tau_0,\vec 0) \in J^-(\scrips )$ if and
only if the whole plane $\{t=\tau_0\;,\;x=0\}$ is included in
$J^-(\scrips )$ (throughout this proof all the causal objects are
taken in $\Mink$). Set $p_i=(\tau_0-\delta,\vec y_i)$, then
$$J^+(K_i)=J^+(p_i)\cap \{t\ge \tau_0\}\;.$$
 Without loss of generality we may assume $\vec
x_1=(a,0,0)$. A simple calculation gives
$$
\pJ ^+(K_1)\cap \pJ ^+(K_2)\cap \pJ
^-(0_y)=\{(\tau_\delta,0,y,z)\;| \ y^2+z^2 =
\tau_\delta^2\}\subset\{x=0\}\;,$$ with
$$\tau_\delta:= \frac {a^2-(|\tau_0|+\delta)^2}{2
(|\tau_0|+\delta)} < 0\;.$$ On the other hand
$$\pJ ^+(p_0)\cap \pJ ^-(0_y)\cap\{x=0\}=\left\{(\frac {\tau_0} 2,0,y,z)\;| \ y^2+z^2 =
\left(\frac {\tau_0} 2\right)^2\right\}\;.$$  It then easily
follows, \emph{e.g.}\/ by symmetry arguments, that $p_0$ will be
in the causal past of $\scrips $ if and only if
$$\frac {\tau_0} 2\le \tau_\delta\;.$$
This last equation is equivalent to \eq{ePc1}. \qed

Proposition~\ref{Pcriterion} together with Remark~\ref{R4.2}
settle the case $I=2$. In order to proceed further, it is
necessary to understand  the geometry of the intersections
$$\pJ^+(p)\cap \pJ^-(0_y)\;,\qquad p\in I^-(0_y)\;.$$
It is convenient to consider general space-time dimensions $n+1$.
Let $\tau_0<0$ and let $p=(\tau_0,\vec q)\in I^-(0)\subset
\R^{n+1}$, with $\vec q \in B(0,|\tau_0|)\subset \R^n$; for the
discussion here all the causal objects are defined with respect to
the Minkowski metric $\eta$ in $\R^{n+1}$. We denote by $\pi$ the
projection along the first, timelike coordinate axis in $\R^{n+1}$
(associated to a coordinate which we denote by $x^0$). We
set\newcommand{\mcOq}{\mcO_{\vec
q}}\newcommand{\pmcOq}{\dot\mcO_{\vec q}}
\beaa \mcU_{\vec q}&:=&J^+((\tau_0,\vec q))\cap \pJ^-(0)\;,  \\
\mcOq&:=& \pi(\mcU_{\vec q})\;.\eeaa A simple computation shows
that the $\mcOq$'s are solid ellipsoids: for $\vec q = (a,\vec
0)$, $|a|<|\tau_0|$, where $\vec 0$ denotes the origin in
$\R^{n-1}$, we have \bel{elli} \mcOa = \Big\{ (x-\frac a2)^2 +
\frac{\rho^2}{1-\frac {a^2}{|\tau_0|^2}} \le
\left(\frac{\tau_0}{2}\right)^2\Big\}\;,\qquad \rho^2:=
(x^2)^2+\ldots+(x^n)^2\;. \ee Here we use the symbol $x$ to denote
the first coordinate in $\R^n$. For further purposes the following
properties of the $\mcOa$'s are useful:
\begin{itemize}
\item The $\mcOa$'s are all cigar shaped, except for the one with
$a=0$ which is a ball of radius $|\tau_0|/2$. \item The $\mcOa$'s
are centred at $(a/2,\vec 0)$, and their extent in the first
coordinate $x$ equals  $\tau_0$, independently of $a$. \item The
intersection of the $\mcOa$'s with the central hyperplane
$\{x=a/2\}$ is an $(n-1)$--dimensional ball of radius
$f(a/|\tau_0|)$, where $f(\beta)=|\tau_0|\sqrt{1-\beta^2}/2$. The
function $f:[0,1]\to[0,|\tau_0|/2]$ is strictly decreasing. \item
We further have \bel{es} (\partial\mcOa)\cap \{x=a/2\}\subset
S^{n-1}(0,|\tau_0|/2)\subset \R^n\;;\ee we have decorated the
sphere $S^{n-1}(0,|\tau_0|/2)$ with a subscript $n-1$ to emphasise
its dimension. The sections \eq{es} are the ``fattest"
$x$--sections of the $\mcOa$'s. Thus, as $a$ increases from $0$ to
$|\tau_0|$ the fattest part of $\mcOa$ thinners, with its boundary
traveling on $S^{n-1}(0,|\tau_0|/2)$ from the equatorial
hyperplane $\{x=0\}$ all the way to the north pole $x=|\tau_0|/2$.
The ellipsoids degenerate to a line in this last limit.
\end{itemize}

The following simple rules complement the above:
\begin{enumerate}
\item Let $\vec x \in B^n(0,|\tau_0|)\setminus \mcOa$. Then the
point $p=(\tau_0,\vec x)$ is in the timelike past of
$$\pJ^-(0)\setminus I^+((\tau_0,a,\vec 0))\;;$$
the required timelike curve is simply a segment of the vertical
line $t\to (t,\vec x)$. \item If $\vec x \in (\partial \mcOa)\cap
\pi(\scrip_\tau)$, then the whole line segment $s(a,\vec 0)
+(1-s)\vec x$, $s\in (0,1)$ is included in $I^-(\scrips)$: Indeed,
by definition of $\mcOa$ there exists a causal geodesic $\Gamma$
from $(a,\vec 0)$ to a point $p$ on $\scrip_\tau$, with $p$
projecting down to $\vec x$ --- the projection $\pi (\Gamma)$ of
$\Gamma$ is the line segment $s(a,\vec 0) +(1-s)\vec x$, $s\in
[0,1]$. The required timelike curve is obtained by a timelike
deformation of the following path: one first moves from
$(\tau_0,\vec x)$ towards the future along the $x^0$ coordinate
line until one meets $\Gamma$, and then one moves along $\Gamma$
until one meets $\scrip_\tau$. We note that if $p$ is not on an
edge of $\scrip_\tau$, then the whole closed segment $s(a,\vec 0)
+(1-s)\vec x$, $s\in [0,1]$, will actually be included in
$I^-(\scrip_\tau)$.
\end{enumerate}

We consider now the case of three or more
$K_i$'s.\begin{figure}[ht]
\begin{center}
  \includegraphics[scale=2.5,width=\textwidth]{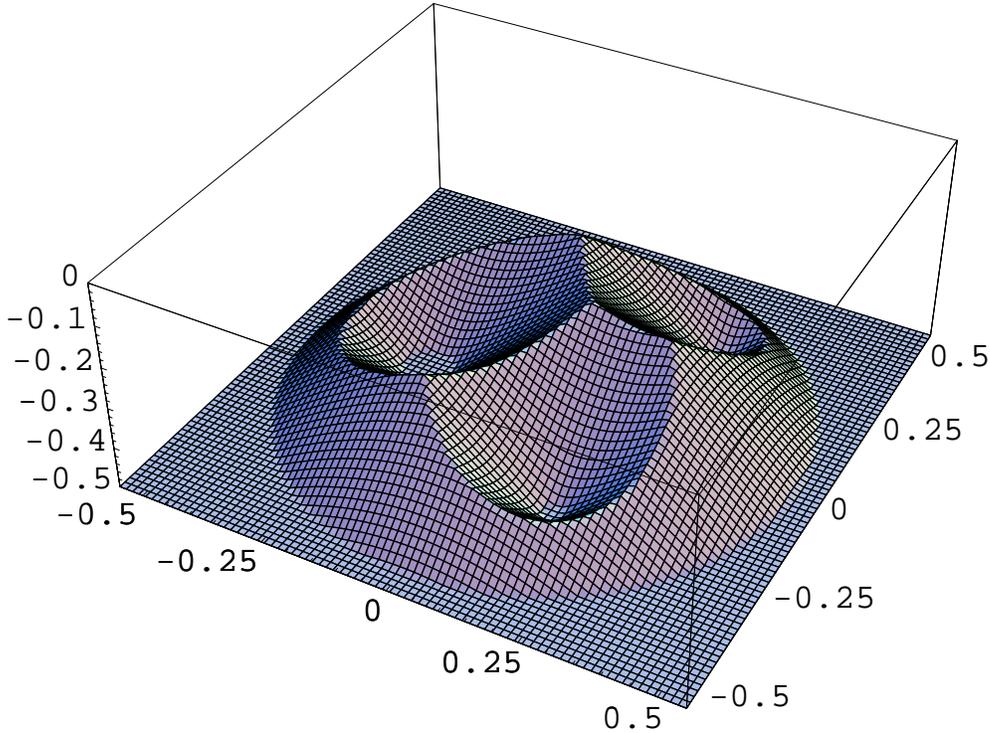}
\caption{\protect A $(2+1)$--dimensional version of the space-time
$\mcM_\tau$, $I=3$, for $\tau$ sufficiently large.}\label{F41a}
\end{center}
\end{figure} The simplest
configuration is that with all the points $\vec y_i$ aligned;
without loss of generality we can then assume that they lie on the
axis $\{y=z=0\}$. We have the following result, where we do not
assume $I=3$:

\begin{Theorem}\label{T4.3}
Suppose that all the $\vec y_i$'s are co-linear, and that the
hypotheses a), b), b') and c) of the beginning of this section
hold. Then for $\eta$'s small enough $\bhzp$ has at least $I$
components. 
\end{Theorem}

\proof The case $I=2$ is covered by Proposition~\ref{Pcriterion},
it is thus sufficient to consider $I\ge 3$. We will construct
hypersurfaces $\mcN_i$ which will be included in $I^-(\scrips)$
for $\delta_i=0$, as all the objects involved depend continuously
upon $\delta_i$ the result for small $\delta_i$'s will follow.

Consider first the case of space--dimension $n=2$, let $\vec
y_i=(a_i,0)$. Then
$$\pi(\scrips)= B(0,|\tau_0|)\setminus \cup_{i}
\mcO_{(a_i,0)}\;.$$ Suppose that $I=3$, without loss of generality
we may assume  $a_1>0$, consider any point $b\in (0,a_1)$. What
has been said concerning the geometry of the $\mcOa$'s implies
that the set \beaa \mcV_b &:=&
\Big\{\mcO_{(b,0)}\cap \pi(\scrips)\Big\}\cap \{x\ge 0\} \\
& = & \Big\{\mcO_{(b,0)}\setminus \{B(0,|\tau_0|/2)\cup
\mcO_{(a_1,0)}\}\Big\}\cap \{x\ge 0\} \eeaa is non-empty, see
Figure~\ref{F2d5}.
 \begin{figure}[t]
\begin{center}
  \include{picture3}
\end{center}
\caption{\protect Three black holes, aligned. The outermost circle
represents the conformal boundary of $\hyp_0$. The dotted region
is the projection on the initial hypersurface $\hyp_0$ of the part
of $\pJ(0_y)$ which has been excised by the removal of
$J^+(K_1)\cup J^+(K_2)$. The shaded region is
$\mcV_b$.}\label{F2d5}\end{figure} Since the dimension is two, the
desired hypersurface $\mcN_1$ is actually a curve, obtained as
follows: let $\vec x$ be any point in $\mcV_b$, let $\gamma_1$ be
the segment $s\vec x + (1-s) (b,0)$, $s\in [0,1]$. Let $\gamma_2$
be the path obtained by first following $\gamma_1$,
and then a line parallel to the $y$ axis. 
By the rules 1) and 2) $\gamma_2$ is included in $I^-(\scrips)$.
We define $\mcN_1$ to be the union of $\gamma_2$ and of its image
under the map $(x,y)\to (x,-y)$. The second hypersurface $\mcN_2$
is obtained by taking the image of $\mcN_1$ under the map
$(x,y)\to (-x,y)$.

\begin{figure}[h]
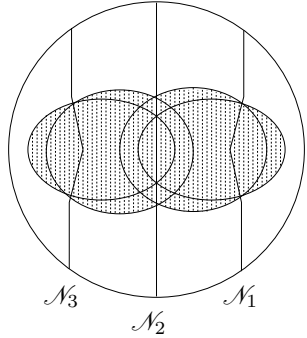

\begin{center}
  \include{picture4}
\end{center}
\caption{\protect Four black holes, aligned. The dotted region is
the projection on the initial hypersurface $\hyp_0$ of the part of
$\pJ(0_y)$ which has been excised by the removal of $J^+(K_1)\cup
J^+(K_2)\cup J^(K_3)$.}\label{F2d4}\end{figure}

It should be obvious to the reader how this construction
generalises to higher $I$'s. We simply note that for $I=2N$ one of
the curves, say $\mcN_N$, can always be chosen to be the axis
$\{x=0\}$. Figure~\ref{F2d4} illustrates the case $I=4$.

 For dimensions $n\ge 3$ the desired hypersurfaces can be
obtained by rotating the curves  constructed above around the axis
$\{y^2=\cdots=y^n=0\}$ using the action of $SO(n-1)\subset SO(n)$.
\qed

Using our discussion of the geometry of the $\mcO_q$'s, with a
little work the reader should be able to verify the following:
\begin{Theorem} Let $|\vec n_i|= 1$ for all $i$, and let $\vec y_i = \lambda \vec n_i$.
There exists $0<\lambda_0<1/2$ such that for all $\lambda_0\le
\lambda<1/2$ the black hole region $\bhzp$ has at least $I$
components.
\end{Theorem}
\qed

One expects that there exist configurations with $I>3$ for which
$\mcE^+_{\tau_+}(0)$ has less than $I$ components. While it is
easy to imagine such configurations, a justification does not seem
to be straightforward.
 \begin{figure}[t]
\begin{center}
  \includegraphics[width=\textwidth]{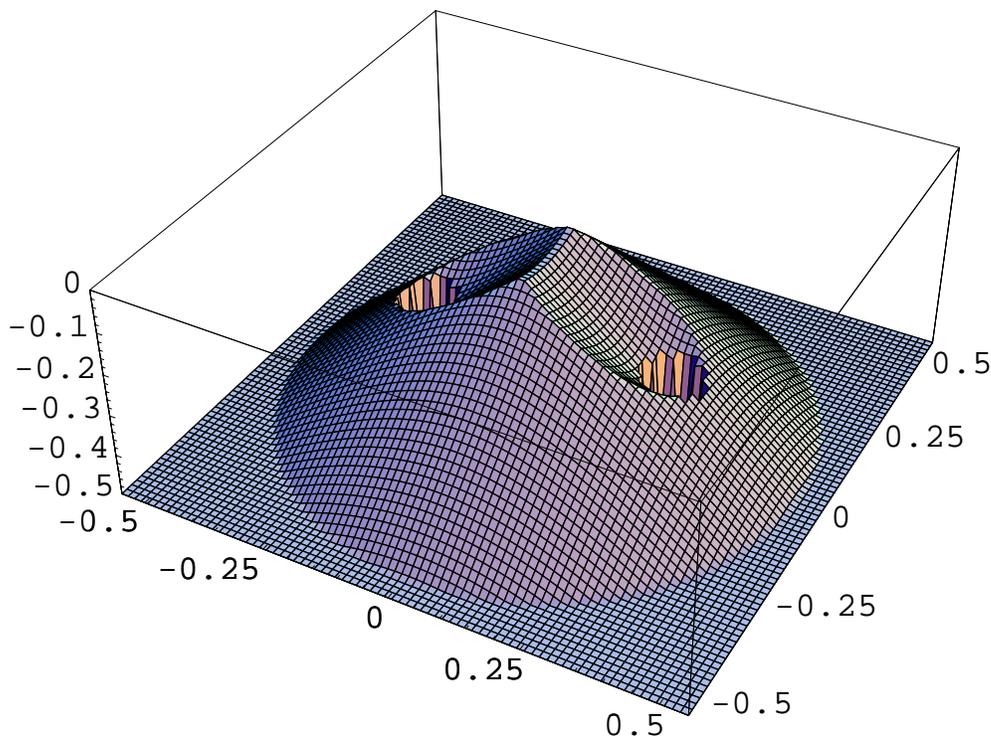}
\end{center}
\caption{Two black holes, with the metric staying close to the
Minkowski one near the spheres $S(\vec x_i,r_i)$ for a significant
time.}\label{F2d4a}
\end{figure}

\begin{figure}[t]
\begin{center}
  \includegraphics[width=\textwidth]{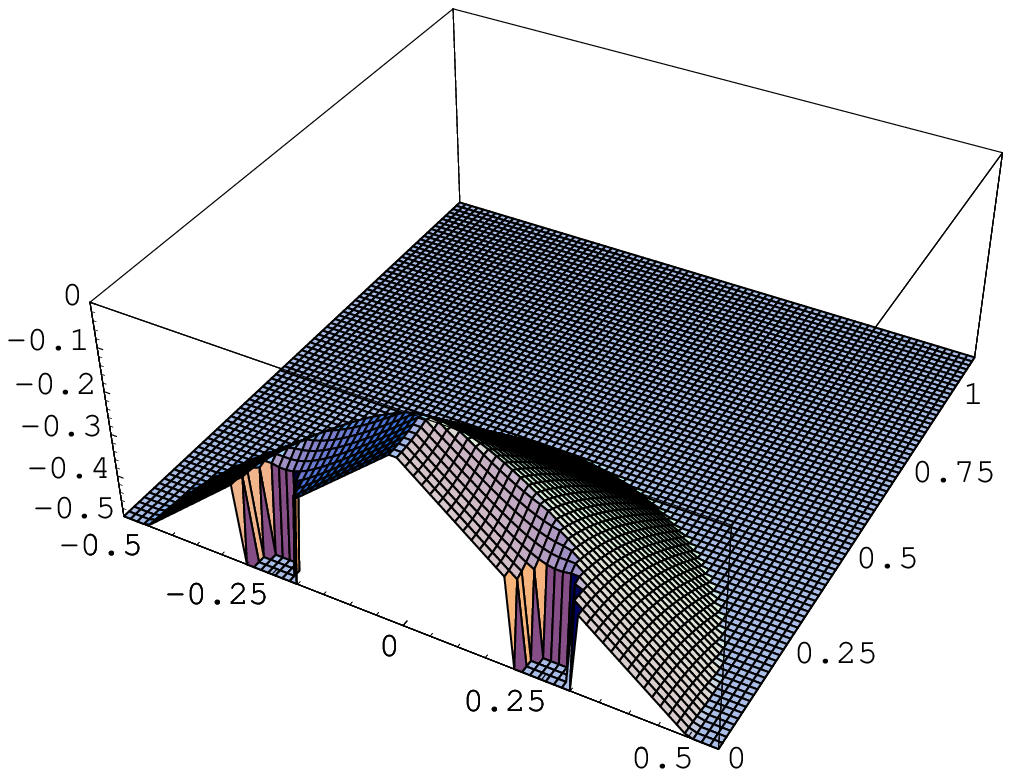}
\end{center}
\caption{The part $y\ge 0$ of the space-times of
Figure~\protect\ref{F2d4a}.}\label{F2d4b}
\end{figure}

So far we have not been assuming anything about how long the
metric remains close to the flat one in a small neighborhood of
the spheres  $S(\vec x_i,r_i)$. The longer this happens, the
larger the set of slices $\{y^0=\tau\}$ at which the horizon has
more than one components. This is made clear by
Figures~\ref{F2d4a}-\ref{F2d4b}.


\acknowledgments PTC acknowledges useful discussions with E.
Delay, G. Galloway and R.~Schoen. He is grateful to the American
Institute of Mathematics, the Stanford University  and the NSF for
financial support at the Stanford Workshop on General Relativity
during part of work on this paper. RM wishes to thank S.~Kerckhoff
for a helpful conversation about $3$-manifold topology. Thanks are
due to J.~Chru\'sciel and S.~Nicolis for help with the
computer-generated drawings.

\begin{figure}[h]
\end{figure}

\bibliographystyle{amsplain}
\bibliography{
../../references/newbiblio,%
../../references/reffile,%
../../references/bibl,%
../../references/Energy,%
../../references/hip_bib,%
../../references/netbiblio}
\end{document}